\documentclass[journal,letterpaper,onecolumn,draft,11pt]{IEEEtran}

\usepackage{amsthm}
\usepackage{amsmath}
\usepackage{amsfonts}
\usepackage{dsfont}
\usepackage{mathrsfs}
\usepackage{pifont}
\usepackage{amssymb}
\usepackage{verbatim}
\usepackage{upgreek}
\usepackage{color}
\usepackage[final]{graphicx}
\usepackage{algorithmic}
\usepackage{algorithm}
\usepackage{epsfig}
\usepackage{eucal}
\usepackage{cite}
\usepackage{subfigure}

\makeatletter
\def\maketag@@@#1{\hbox{\m@th\normalfont\normalsize#1}}
\makeatother

\newcommand{\bbm}{\begin{bmatrix}}
\newcommand{\ebm}{\end{bmatrix}}

\newcommand{\bit}{\begin{itemize}}
\newcommand{\eit}{\end{itemize}}

\newcommand{\ben}{\begin{enumerate}}
\newcommand{\een}{\end{enumerate}}

\newcommand{\bdesc}{\begin{description}}
\newcommand{\edesc}{\end{description}}

\newcommand{\bea}{\begin{array}}
\newcommand{\eea}{\end{array}}

\newcommand{\tr}{\mbox{\rm Tr}\, }

\newcommand{\beqa}{\begin{eqnarray}}
\newcommand{\eeqa}{\end{eqnarray}}

\newcommand{\ds}{\displaystyle}

\newcommand{\Comment}[1]{}

\def\N{{\mathds N}}

\def\R{{\mathds R}}

\def\C{{\mathds C}}

\def\cC{\mbox{$\CMcal C$}}

\def\cL{\mbox{$\mathcal L$}}
\def\cN{\mbox{$\CMcal N$}}

\def\cX{\mbox{$\mathcal X$}}
\def\cY{\mbox{$\mathcal Y$}}

\newcommand{\be}{\begin{equation}}
\newcommand{\bbq}{\bar{q}}
\newcommand{\ee}{\end{equation}}

\newcommand{\bzero}{{\mbox{\boldmath $0$}}}

\newcommand{\boa}{{\mbox{\boldmath $a$}}}

\newcommand{\bd}{{\mbox{\boldmath $d$}}}
\newcommand{\boe}{{\mbox{\boldmath $e$}}}

\newcommand{\bm}{{\mbox{\boldmath $m$}}}

\newcommand{\bt}{{\mbox{\boldmath $t$}}}

\newcommand{\bv}{{\mbox{\boldmath $v$}}}
\newcommand{\bx}{{\mbox{\boldmath $x$}}}

\newcommand{\bz}{{\mbox{\boldmath $z$}}}

\newcommand{\bA}{{\mbox{\boldmath $A$}}}
\newcommand{\bB}{{\mbox{\boldmath $B$}}}
\newcommand{\bC}{{\mbox{\boldmath $C$}}}
\newcommand{\bD}{{\mbox{\boldmath $D$}}}

\newcommand{\bH}{{\mbox{\boldmath $H$}}}
\newcommand{\bI}{{\mbox{\boldmath $I$}}}

\newcommand{\bM}{{\mbox{\boldmath $M$}}}

\newcommand{\bP}{{\mbox{\boldmath $P$}}}

\newcommand{\bS}{{\mbox{\boldmath $S$}}}

\newcommand{\bU}{{\mbox{\boldmath $U$}}}
\newcommand{\bV}{{\mbox{\boldmath $V$}}}

\newcommand{\bZ}{{\mbox{\boldmath $Z$}}}

\newcommand{\diag}{\mbox{diag}\, }

\newcommand{\bLambda}{{\mbox{\boldmath $\Lambda$}}}

\newcommand{\dmax}{\begin{displaystyle}\max\end{displaystyle}}
\newcommand{\dmin}{\begin{displaystyle}\min\end{displaystyle}}

\newcommand{\test}{\mbox{$
\begin{array}{c}
\stackrel{ \stackrel{\textstyle H_1}{\textstyle >} }{
\stackrel{\textstyle <}{\textstyle H_0} }
\end{array}
$}}

\title{A Sparse Learning Approach to the Detection of Multiple Noise-Like Jammers}

\vspace{0.1cm}

\author{Linjie Yan, Pia Addabbo, \IEEEmembership{Senior Member, IEEE}, Yuxuan Zhang,
Chengpeng Hao, \IEEEmembership{Senior Member, IEEE}, Jun Liu, \IEEEmembership{Senior Member, IEEE},
Jian Li, \IEEEmembership{Fellow, IEEE}, and Danilo Orlando, \IEEEmembership{Senior Member, IEEE}
\thanks{Linjie Yan, Yuxuan Zhang, and Chengpeng Hao are with Institute of Acoustics, Chinese 
Academy of Sciences, Beijing, China.
E-mail: {\tt yanlinjie16@163.com,zhangyuxuan@mail.ioa.ac.cn,haochengp@mail.ioa.ac.cn}.} 
\thanks{Pia Addabbo is with Universit\`a degli studi Giustino Fortunato, Benevento, Italy. E-mail: {\tt 
p.addabbo@unifortunato.eu}.} 
\thanks{Jun Liu is with the Department of Electronic Engineering and Information Science, University of Science and 
Technology of China, Hefei 230027, China. E-mail: {\tt junliu@ustc.edu.cn}.}
\thanks{Jian Li is with the Department of Electrical and Computer Engineering, University of Florida, Gainesville, 
FL 32611, USA. E-mail: {\tt li@dsp.ufl.edu}.}
\thanks{Danilo Orlando is with the Engineering Faculty of Universit\`a degli Studi ``Niccol\`o Cusano'', 
via Don Carlo Gnocchi 3, 00166 Roma, Italy. E-mail: {\tt danilo.orlando@unicusano.it}.}
}

\begin{document}

\maketitle

\begin{abstract}
In this paper, we address the problem of detecting multiple Noise-Like Jammers (NLJs) through 
a radar system equipped with an array of sensors. 
To this end, we develop an elegant and systematic framework wherein two architectures are 
devised to jointly detect an unknown number of NLJs
and to estimate their respective angles of arrival. The followed approach
relies on the likelihood ratio test in conjunction with 
a cyclic estimation procedure which incorporates at the design stage a sparsity promoting prior.
As a matter of fact, the problem at hand owns an inherent sparse nature which is suitably exploited.
This methodological choice is dictated by the fact that, from a mathematical point of view, 
classical maximum likelihood approach leads to intractable optimization problems (at least to the best
of authors' knowledge) and, hence, a suboptimum approach represents a viable means to solve them. 
Performance analysis is conducted on simulated data and shows the effectiveness
of the proposed architectures in drawing a reliable picture of 
the electromagnetic threats illuminating the radar system.
\end{abstract}


\section{Introduction}
\label{Sec:Introduction}
In the last decades, the radar art has made great strides due to the advances in technology. In fact, the last-generation 
processing boards are capable of performing huge amounts of computations in a very short time leading to flexible fully-digital architectures.
In addition, this abundance of computation power has allowed for the development of radar systems endowed with more and more sophisticated processing schemes.
A tangible example is represented by search radars which are primarily concerned with the detection of targets 
buried in thermal noise, clutter, and, possibly, intentional interference, also known
as Electronic Countermeasure (ECM) \cite{Richards,antennaBased,EW101,ScheerMelvin}. In this context, 
the open literature is continuously enriched with novel contributions 
that lead to enhanced performances at the price of an increased computational load
\cite{kelly1986adaptive,robey1992cfar,gini1,RicciRao,WLiuRao, Yuri01, BOR-Morgan, Liu1, Liu2, LiuSun19, CP00, 
DeMaioInvPersymmetry, DeMaioSymmetric, fogliaPHE_SS, HaoSP_HE}. Another example related to the potentialities provided by fully-digital architectures 
is connected with Adaptive Digital BeamForming (ADBF) techniques \cite{ScheerMelvin,antennaBased}, since
they can suitably combine digital samples at the output of each channel according to the specific requirement. 
Remarkably, by means of ADBF techniques, the transmit/receive antenna beam patterns can be suitably shaped 
preventing the system 
engineer from the duplication of hardware resources.
For instance, ADBF can be used to build up the auxiliary beam used by the 
{\em SideLobe Blanker} (SLB) \cite{antennaBased,FarinaGini,DeMaioFarinaGini,AubryDeMaioFarina,PiezzoDeMaioFarina} 
exploiting the entire array without the need of additional antennas.
The SLB is an Electronic Counter-CounterMeasure (ECCM) against pulsed intentional interferences (or coherent jammers) entering the antenna sidelobes, 
which, in turn, are ECMs. Note that ECCM techniques
can be categorized as antenna-related, transmitter-related, receiver-related, and signal-processing-related 
depending on the main radar subsystem where they take place \cite{FarinaSkolnik}.

Besides coherent jammers, any radar might also be a victim of noise-like interfering 
signals, also referred to as
Noise-Like Jammers (NLJs),
by an adversary force. This electronic attack is aimed at preventing detection or denying accurate 
measurement of target information (Doppler and/or Range) \cite{ScheerMelvin}
by generating nondeceptive interference which blends into the thermal noise of the radar receiver. 
As a consequence, the radar sensitivity is degraded due to the increase of the
constant false alarm rate threshold which adapts to the higher level of noise \cite{antennaBased,ScheerMelvin}. In addition,
this increase makes more difficult to know that jamming is taking place \cite{EW101,FarinaSkolnik}. 
Under the NLJ attack, the SLB becomes ineffective since it would inhibit the detection of true targets for most of the time.
In these situations, the Sidelobe Canceler (SLC) represents a viable ECCM \cite{antennaBased,FarinaSLC,Reed}. As a matter of fact, it 
exploits an additional auxiliary\footnote{Note that a system with sidelobe canceling capabilities 
is equipped with both the main antenna array devoted to target detection and an auxiliary 
array used to cancel the NLJs.} 
array of antennas (with suitable gains) to adaptively estimate the NLJ Angle of Arrival (AoA)
and places nulls in the sidelobes of the main receiver beam along the estimated AoA. 
In a fully-digital architecture, the task of the SLC can be accomplished by applying ADBF techniques without the 
use of additional hardware (signal-processing-related ECCM). 

However, the application of ADBF techniques might increase the computational burden of the signal processing unit since they require
the computation and the inversion of a sample covariance matrix in addition to possible AoA estimation. 
These operations consume hardware resources which are shared
among the different radar functions and, due to the restrictive requirements on radar reaction time, 
they cannot occur at every dwell regardless whether or not NLJs are illuminating the radar. 
Thus, it would be highly desirable a preliminary stage capable
of detecting NLJs and, possibly, estimating the relevant NLJ parameters. Once the presence of NLJs is declared,
the estimated parameters are used by ADBF techniques to contrast the interfering actions.
Following this reasoning, in \cite{jammerDetection}, the authors develop a decision scheme which decides for the presence of one NLJ by comparing 
the spectral properties of reference cells, not affected by jammer returns, 
with those of Cells Under Test (CUT); no additional information about the NLJ is provided.
The case of multiple NLJs is addressed in \cite{CarotenutoNLJ}, where the original binary hypothesis 
test is transformed into a multiple-hypothesis problem and the Model Order Selection (MOS) rules \cite{StoicaBabu1,Stoica1,Kay1,Kay2,kay2005multifamily,WaxKailath}
are exploited to conceive two-stage detection architectures, where the first stage
provides an estimate of the active NLJs number under the constraint of an upper bound to it, while
the second stage is devoted to the detection of the estimated number of NLJs allowing for 
the control of the false jammer detection probability.
However, these two-stage architectures are not capable
of providing any information about either the AoA or the received power of the detected NLJs.

\begin{figure}[tb!]
\begin{center}
\includegraphics[scale=0.5]{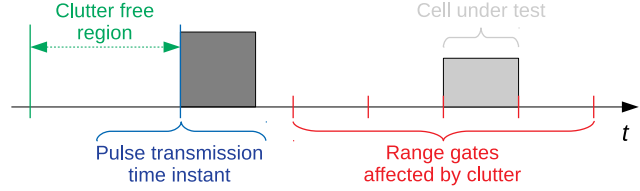}
\caption{Acquisition procedure of clutter free data for spatial processing.}
\label{fig:clutterFreeData}
\end{center}
\end{figure}

\begin{figure}[tb!]
\begin{center}
\includegraphics[scale=0.5]{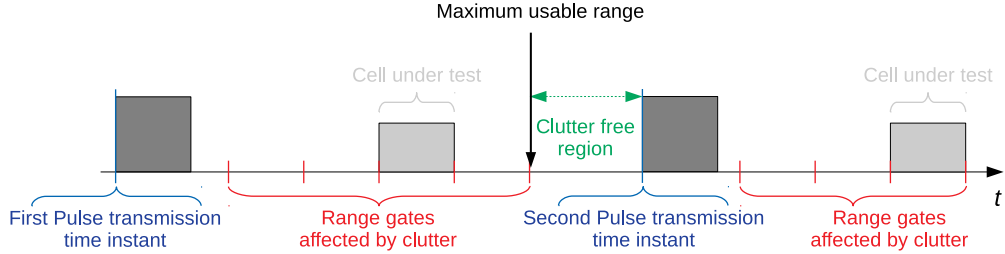}
\caption{Acquisition procedure of clutter free data for temporal processing.}
\label{fig:clutterFreeData2}
\end{center}
\end{figure}

With the above remarks in mind, in this paper, we address 
the same detection problem as in \cite{CarotenutoNLJ} by developing an elegant and systematic
framework for the joint detection 
of multiple NLJs and the estimation of the respective relevant parameters, which include the AoAs 
and the number of threats\footnote{Recall that in \cite{CarotenutoNLJ} the focus is limited 
to the interference subspace detection \cite{VanTrees4} without providing any side information.}.
To  this end, we assume that a set of data free of clutter components 
and affected by thermal noise and possible NLJ components \cite{antennaBased,ward,DoppioTraining}
is available at the receiver. As a matter of fact, it can be collected 
by noticing that the clutter contribution is,
in general, range-dependent and tied up to the transmitted waveform. Therefore, it is
possible to acquire data free of clutter components and affected by the thermal noise
and possible jamming signals only. For instance, for a system employing pulse-to-pulse
frequency agility which transmits one pulse, clutter-free data can be collected before
transmitting the pulse waveform by listening to the environment (see Figure \ref{fig:clutterFreeData}).
Another example of practical interest concerns radar systems transmitting coherent pulse trains with 
a sufficiently high pulse repetition interval. In this case, data collected before transmitting
the next pulse and at high ranges (or after the instrumental range), result free of clutter
contribution (see Figure \ref{fig:clutterFreeData2}). 
Now, under these assumptions, 
the newly proposed framework exploits a sparse representation of the problem at hand and
resorts to suitable cyclic optimization procedures \cite{Stoica_alternating} to devise two 
architectures where the AoA and power estimation is concurrent with the detection without
any subsequent estimation stage or constraint on the number of NLJs. 
Following the lead of \cite{slim}, we assume that NLJ parameters are random and obey a 
prior that promotes sparsity. However, the latter is conceived for the specific case at hand giving rise
to a new optimization problem and, hence, new analytical derivations.
Remarkably, the considered sparsity-based estimation allows for an 
increase of the angular resolution (at least for high NLJ powers as shown in Section \ref{Sec:Performance}).
Finally, the obtained estimates are plugged into a Likelihood Ratio Test (LRT) aimed at detecting
the presence of NLJs. 
The above aspects represent the main technical contribution of this work and, at least to the best of 
authors' knowledge, appear for the first time in this paper. 

It is also important to underline that these methodology choices lead to suboptimum 
solutions which are dictated by the fact that the plain Maximum Likelihood Approach (MLA) exhibits a difficult mathematical tractability.
Performance analysis, conducted on simulated data, 
points out the effectiveness of the newly proposed decision schemes from the point of view of
both detection and estimation capabilities also in comparison with their natural competitors.

The remainder of the paper is organized as follows. Section \ref{Sec:Problem_Formulation} 
is devoted to problem formulation and definition of quantities used in the next derivations, 
while the design of the detection architectures and the estimation procedures are described in
Section \ref{Sec:Architecture_Designs}. Section \ref{Sec:Performance} shows 
the effectiveness of the proposed strategies through numerical examples on simulated data. 
Finally, Section \ref{Sec:Conclusions} contains concluding remarks and charts a course for future works; 
some mathematical derivations and proofs are confined to the appendices.

\subsection{Notation}
In the sequel, vectors and matrices are denoted by boldface lower-case and upper-case letters, respectively.
The $i$th entry of a vector $\boa$ is represented by $\boa(i)$ whereas
symbols $\det(\cdot)$, $\tr(\cdot)$, $(\cdot)^T$, and $(\cdot)^\dag$ denote the determinant, trace, transpose, 
and conjugate transpose, respectively.
Symbol $\|\cdot\|$ denotes the Euclidean norm of a vector. 
As to numerical sets, $\N$ is the set of natural numbers, $\R$ is the set of real numbers, $\R^{N\times M}$ is the Euclidean space of $(N\times M)$-dimensional 
real matrices (or vectors if $M=1$), $\R_+^{N\times M}$ is the set of $(N\times M)$-dimensional 
real matrices (or vectors if $M=1$) whose entries are greater than or equal to zero, $\C$ is the set of 
complex numbers, and $\C^{N\times M}$ is the Euclidean space of $(N\times M)$-dimensional 
complex matrices (or vectors if $M=1$). The modulus of a real number $x$ is denoted by $|x|$.
$\bI$ and $\bzero$ stand for the identity matrix and the null vector or matrix of proper size. 
Symbol $\propto$ means that the left-hand side is proportional to the right-hand side.
Given a vector $\boa\in\C^{N\times 1}$, $\diag(\boa)\in\C^{N\times N}$ indicates 
the diagonal matrix whose $i$th diagonal element is the $i$th entry of $\boa$.
The acronym pdf stands for probability density function and the conditional pdf of a random variable $x$ given 
another random variable $y$ is denoted by $f(x|y)$. Finally, we write $\bx\sim\cC\cN_N(\bm, \bM)$ if $\bx$ is a 
complex circular $N$-dimensional normal vector with mean $\bm$ and positive definite covariance matrix $\bM$.

\section{Problem Formulation and Preliminary Definitions}
\label{Sec:Problem_Formulation}
Consider a radar system equipped with $N\geq 2$ spatial channels which is listening to the environment. 
The incoming signal is firstly conditioned by means of a baseband 
down-conversion, then, it is pre-processed and properly sampled. The samples are, then, organized to form $N$-dimensional vectors 
denoted by $\bz_k$, $k=1,\ldots,K$, with $K\geq N_j$ being the total number of
listening data and $N_j \leq N$ the number of NLJs. The detection problem at hand can be formulated as
\be
\left\{
\begin{array}{lll}
H_0: & \bz_k \sim \cC\cN_N(\bzero, \bM_0), & \! k = 1, \ldots, K,
\\
H_1: & \bz_k \sim \ds\cC\cN_N\left(\bzero, \bM_1\right), & \! k = 1, \ldots, K,
\end{array}
\right.
\label{eqn:decisionProblem}
\ee
where $\bM_0 = \sigma^2_n\bI$ and
\be
\bM_1 = \sigma^2_n\bI + \sum_{i=1}^{N_j}d_i \bv(\bar{\theta}_{i})\bv(\bar{\theta}_{i})^\dag.
\label{eqn:model_ini}
\ee
In the last equations, $\sigma^2_n\geq 1$ and\footnote{As explained in Appendix 
\ref{App:prior}, the lower bound on the thermal noise power is required to ensure a {\em good behavior} 
for the prior associated to $d_i$ that will be introduced in the next section. From a practical point of view, 
this lower bound can be handled by exploiting a suitable numerical representation used by the signal 
processing unit.} $d_i>0$ are the powers of thermal noise 
and the $i$th jammer, respectively,
$\bar{\theta}_{i}$ is the AoA of the $i$th jammer measured with respect to the array broadside, and
$\bv(\theta)$ is the array steering vector pointed along $\theta$ whose expression is
$\bv(\theta)=\frac{1}{\sqrt{N}}\left[1, e^{j 2\pi (d/\lambda) \sin(\theta)},\ldots,e^{j 2\pi (d/\lambda)(N-1)\sin(\theta)} \right]^T$
with $d$ the array interelement spacing and $\lambda$ the carrier wavelength. Moreover, under each hypothesis, $\bz_k$s are statistically independent.

\begin{figure}[tb!]
\begin{center}
\includegraphics[scale=0.5]{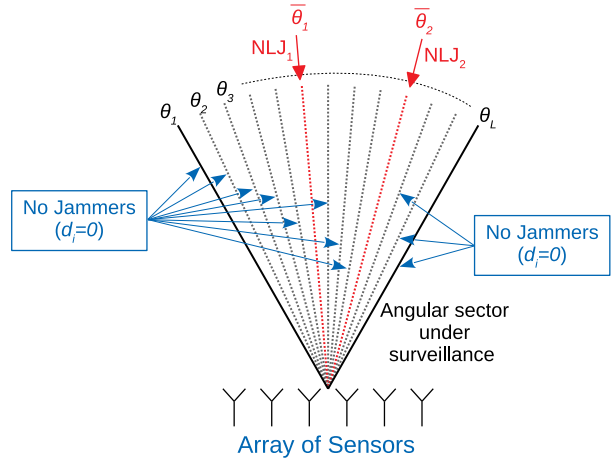}
\caption{A pictorial representation of the hidden sparse nature of model \eqref{eqn:model_ini}
assuming $N_j=2\ll L$.}
\label{fig:sparseNature}
\end{center}
\end{figure}

In order to bring to light the sparse nature of the problem, let us sample the angular sector under surveillance to form a discrete and 
finite set of angles denoted by $\Theta=\left\{ \theta_1, \ldots, \theta_L \right\}$ with $L\gg N_j$ and $\theta_1\leq\ldots\leq\theta_L$. 
In addition, we assume that $\forall i=1,\ldots,N_j$, $\bar{\theta}_i\in\Theta$. Thus, if we define 
a vector $\bd=[d_1,\ldots,d_L]^T\in \mathbb{\R}_+^{L\times 1}$ such that
\be
\forall k=1,\ldots,L \ : \ 
\begin{cases}
d_k > 0, & \mbox{if } \theta_k=\bar{\theta}_i,
\\
d_k = 0, & \mbox{otherwise},
\end{cases}
\ee
it follows that $\bd$ is sparse (since $L\gg N_j$) and the ICM under $H_1$ can be recast as
\be
\bM_1 = \sigma^2_n\bI + \bV\bD\bV^\dag,
\label{eqn:covarianceModel}
\ee
where $\bV=[\bv(\theta_1),\ldots,\bv(\theta_L)]$ is the dictionary and $\bD=\diag(\bd)$. 
In Figure \ref{fig:sparseNature}, we show a pictorial representation of the hidden sparse 
nature of \eqref{eqn:model_ini}.
Thus, the formal structure of the 
detection problem at hand can be expressed in terms of the sparse vector
$\bd$ as follows
\be
\left\{
\begin{array}{lll}
H_0: & \bd = \bzero,
\\
H_1: & \bd \neq \bzero \ \mbox{(with nonnegative entries)}.
\end{array}
\right.
\label{eqn:decisionProblemSparse}
\ee
Finally, we conclude this section by providing the expression of the pdf of $\bZ=[\bz_1,\ldots,\bz_K]$ 
under $H_i$, $i=0,1$, which will be used in the next developments, namely
\be
f_i(\bZ; \sigma_n^2, i\bd, H_i)=
\left[
\frac{1}{\pi^N \det(\sigma^2_n\bI + i\bV\bD\bV^\dag)}
\right]^K 
\exp\left\{ -\tr\left[(\sigma^2_n\bI + i\bV\bD\bV^\dag)^{-1}\bZ\bZ^\dag\right] \right\}.
\label{eqn:pdf_Z}
\ee

\section{Architecture Designs}
\label{Sec:Architecture_Designs}
As stated in Section \ref{Sec:Introduction}, the MLA for this problem leads to intractable mathematics and, hence,
we resort to a suboptimum iterative approach. With this remark in mind, in this section, we derive two decision schemes for problem \eqref{eqn:decisionProblemSparse} which differ in the adaptivity with respect to the
thermal noise power. Specifically, the former estimates $\bd$ assuming that $\sigma^2_n$ is known and, then, 
replaces it with an estimate
which is assumed available at the receiver (a point better explained in the next subsection). 
The latter jointly estimates $\bd$ and $\sigma^2_n$ by means of a cyclic optimization procedure.
In both cases, the structure of the decision statistic is given by the likelihood ratio and the decision rule is given by the LRT, whose expression is
\be
\Lambda(\bZ;\bd,\sigma_n^2)=\frac{f_1(\bZ; \sigma_n^2, \bd, H_1)}{f_0(\bZ; \sigma_n^2, \bzero, H_0)}\test\eta,
\ee
where $\eta$ is threshold\footnote{Hereafter, we denote by $\eta$ any modification of the detection threshold.} 
to be set in order to guarantee the required Probability of False Jammer Detection ($P_{fjd}$).

\subsection{Adaptive detector for unknown $\bd$}
\label{Subsec:knownSigma}
Let us assume that $\sigma_n^2$ is known and, following the lead of \cite{slim}, that the entries 
of $\bd$ are jointly distributed according to a sparsity promoting (possibly improper) prior 
given by\footnote{There does not exist a specific criterion to 
select the prior for the considered framework. Nevertheless, the choice of this prior
raises from an analysis of the achievable performance.}
\be
f_d(\bd;\sigma^2_n,q) \propto \frac{\ds[\det(\sigma_n^2\bI + \bV\bD\bV^\dag)]^{K-1}}
{\ds\prod_{i=1}^L \exp\left\{\frac{K}{q}(d_i^q-1)\right\}}
\label{eqn:prior}
\ee
with a (possible) positive constant of proportionality, where $q\in\Omega_q=(0,\ 1]$ is a parameter allowing 
for sparsity control. In Appendix \ref{App:prior}, we investigate the behavior of $f_d(\bd;\sigma^2_n,q)$
with respect to $\bd$ and $q$.
Thus, the logarithm of the joint pdf of $\bZ$ and $\bd$ under $H_1$ can be written as
\begin{align}
\log f(\bZ,\bd;\sigma^2_n,q)=&\log f(\bZ|\bd;\sigma^2_n,q) + \log f_d(\bd;\sigma^2_n,q) \nonumber
\\
\approx &-KN\log\pi-\log\det(\sigma^2_n\bI + \bV\bD\bV^\dag)
-\tr\left[(\sigma^2_n\bI + \bV\bD\bV^\dag)^{-1}\bS\right] \nonumber
\\
&-\sum_{i=1}^L \frac{K}{q}(d_i^q-1)=g(\bd;\sigma_n^2,q),
\end{align}
where $\bS=\bZ\bZ^\dag$ and proportionality constant of the prior of $\bd$ has been neglected.
Now, we proceed by setting to zero the first derivative of $g(\bd;\sigma_n^2,q)$ with respect to $d_i$ \cite{VanTrees4}, namely
$\frac{\partial}{\partial d_i}\left[g(\bd;\sigma_n^2,q)\right] = 0$, $i=1,\ldots,L$,
which leads to the following equations
\begin{align}
& -\tr\left[(\sigma^2_n\bI + \bV\bD\bV^\dag)^{-1}\bv(\theta_i)\bv(\theta_i)^\dag\right]
+\tr\left[ (\sigma^2_n\bI + \bV\bD\bV^\dag)^{-1}\bS (\sigma^2_n\bI + \bV\bD\bV^\dag)^{-1}\bv(\theta_i)\bv(\theta_i)^\dag \right] \nonumber
\\
& - K \frac{d_i}{d_i^{2-q}} = 0 \Rightarrow 
d_i = 
\begin{cases}
\frac{d_i^{2-q}}{K} \bv(\theta_i)^\dag \bH(\bd) \bv(\theta_i), \ \mbox{if} \  \bv(\theta_i)^\dag \bH(\bd) \bv(\theta_i)>0,
\\
0, \quad \mbox{otherwise},
\end{cases}
\quad i=1,\ldots,L.
\label{eqn:fixedPointD}
\end{align}
Observe that when $K \gg N$ or $K>N \gg 0$, 
then $\bS\approx K (\sigma^2_n\bI + \bV\bD\bV^\dag)$ and, hence, the following matrix
\be
\bH(\bd)=\left[(\sigma^2_n\bI + \bV\bD\bV^\dag)^{-1}\bS (\sigma^2_n\bI + \bV\bD\bV^\dag)^{-1} - (\sigma^2_n\bI + \bV\bD\bV^\dag)^{-1}
\right]
\label{eqn:Hdefinition}
\ee
is positive definite. Equations \eqref{eqn:fixedPointD} can be written in matrix form as
\be
\bd=\frac{\bP_q}{K}
\begin{bmatrix}
\max\{\bv(\theta_1)^\dag \bH(\bd) \bv(\theta_1),0\}
\\
\vdots
\\
\max\{\bv(\theta_L)^\dag \bH(\bd) \bv(\theta_L),0\}
\end{bmatrix}
\ee
with $\bP_q=\diag(d_1^{2-q},\ldots,d_L^{2-q})$ and $\max\{\cdot,0\}$ is used to guarantee the constraint that the entries of $\bd$ are
nonnegative. Now, given a preassigned value for $q$, let us 
assume that an initial estimate of $\bd$, denoted by ${\bd}^{(0)}_q$, 
is available, then, it is possible to apply a cyclic optimization \cite{slim,Stoica_alternating} whose $n$th step is given by
\be
{\bd}_q^{(n+1)}=\frac{\bP_q^{(n)}}{K}
\begin{bmatrix}
\max\left\{\bv(\theta_1)^\dag \bH\left({\bd}_q^{(n)}\right) \bv(\theta_1),0\right\}
\\
\vdots
\\
\max\left\{\bv(\theta_L)^\dag \bH\left({\bd}_q^{(n)}\right) \bv(\theta_L),0\right\}
\end{bmatrix}.
\label{eqn:iterative_d}
\ee
It is important to highlight that the described procedure leads to a nondecreasing sequence of values 
for the cost function $g(\bx;\sigma_n^2,q)$, $\bx\in\R^{L\times 1}_+$. As a matter of fact, first note that
$g(\bx;\sigma^2_n,q)$ is continuous and
\be
\begin{cases}
\ds\lim_{\| {\bf x} \|\rightarrow 0} g(\bx;\sigma^2_n,q) = C < 0,
\\
\ds\lim_{\| {\bf x}\|\rightarrow +\infty} g(\bx;\sigma^2_n,q) = -\infty.
\end{cases}
\ee
The above conditions imply that $g(\bx;\sigma^2_n,q)$ is upper bounded over $\R^{L\times 1}_+$. Moreover, exploiting
{\em Lemma 1} and {\em Theorem 2} of \cite{slim} it is not difficult to show that
\be
g(\bd^{(n)}_q;\sigma^2_n,q)\leq g(\bd^{(n+1)}_q;\sigma^2_n,q).
\ee
It still remains to estimate $q$. To this end, let us sample $\Omega_q$ to come up with a 
finite set of admissible values for $q$ denoted by $\bar{\Omega}_q$.
Now, given $q\in\bar{\Omega}_q$ and the maximum number of jammers $N_{j,\max}$, 
let us denote the number of peaks by $h(q)$ ($\leq N_{j,\max}$), in ${\bd}_q^{(n+1)}$ as follows
\begin{enumerate}
\item sort the entries of ${\bd}_q^{(n+1)}$ from the largest to the smallest and form vector $\tilde{\bd}_{q}$;
\item select $h(q)$ returning the lowest value of
\be
\mbox{BIC}_q = 2K\log\det(\sigma^2_n\bI + \bV\bD_q\bV^\dag) + 2\tr\left[(\sigma^2_n\bI + \bV\bD_q\bV^\dag)^{-1}\bS\right] + h(q) \log\left(2NK\right),
\label{eqn:BICq}
\ee
with\footnote{Note that \eqref{eqn:BICq} is reminiscent of the Minimum Description Length criterion applied in
\cite{WaxKailath}.} $\bD_q=\diag(\widehat{\bd}_q)$ and $\widehat{\bd}_q$ being computed as described 
in Appendix \ref{App:compute_d} (and summarized in Algorithm \ref{Alg:refinement_d}), where
an alternating optimization procedure is applied by setting to zero the entries of $\bd$ whose indices are different 
from those of $\{ \tilde{\bd}_{q}(1),\ldots, \tilde{\bd}_{q}(h(q)) \}$ computed with respect to the element indices of ${\bd}_q^{(n+1)}$.
\end{enumerate}
As a result, we obtain the set $\{\mbox{BIC}_q: q\in\bar{\Omega}_q\}$ and the estimate of $q$ is obtained as
\be
\widehat{q} = \arg\min_{q\in\bar{\Omega}_q} \mbox{BIC}_q.
\ee
Finally, several stopping criteria can be thought to interrupt the cycles. For instance, they can rely on a maximum number of iterations or
on the relative variations with respect to the values returned at the previous iteration. The entire procedure is outlined in Algorithm \ref{Alg:SLIM_known_sigma}.

\begin{algorithm}[tbp!]
\caption{Cyclic algorithm to refine the estimate of $\bd$}
\label{Alg:refinement_d}
\begin{algorithmic}[1]
	\REQUIRE  ${\bd}_q^{(n+1)}$, $q \in (0,1]$, $\bS$, $\bV$, $\sigma^2_n$, and $h(q)$.
	\ENSURE $\widehat{\bd}_q$.
	\STATE Set $n=0$ and $\bar{\bd}_q^{(0)}$ is obtained by setting to zero the entries of ${\bd}_q^{(n+1)}$ different from the first $h(q)$ peaks.
	\STATE Set $n=n+1$ and $i=0$.
	\STATE Set $i=i+1$.
	\STATE Compute $\bar{\Omega}=\{k\in\N: \ \bar{\bd}^{(n-1)}_q(k) > 0\}$, $\bA_{1:i}^{(n-1)} = \sigma^2_n\bI + \sum_{k\in\bar{\Omega}\setminus \Omega_{1:i}}\bar{\bd}^{(n-1)}_q(k)\bv(\theta(k))\bv(\theta(k))^\dag
+\bC^{(n)}_i$ with $\bC^{(n)}_i=\ds\sum_{h\in\Omega_{1:i}\setminus \{i\}}\bar{\bd}^{(n)}_q(h)\bv(\theta(h))\bv(\theta(h))^\dag$ and $\Omega_{1:i}=\{k\in\bar{\Omega}: k\leq i\}$
	\STATE Compute 
	\[
	\bar{\bd}^{(n)}_q(i) = \max\left\{\frac{\ds\bv(\theta(i))^\dag \left[\bA_{1:i}^{(n-1)}\right]^{-1} \bS \left[\bA_{1:i}^{(n-1)}\right]^{-1} \bv(\theta(i))
-K\bv(\theta(i))^\dag \left[\bA_{1:i}^{(n-1)}\right]^{-1} \bv(\theta(i))}
{\ds K\left\{\bv(\theta(i))^\dag \left[\bA_{1:i}^{(n-1)}\right]^{-1} \bv(\theta(i))\right\}^2},0\right\}
	\].
	\STATE If $i<L$ go to step $3$ else go to step $7$.
	\STATE If the stopping condition on $n$ is satisfied go to step $8$ else go to $2$.
	\STATE Return $\widehat{\bd}_q=\bar{\bd}_q^{(n)}$.
\end{algorithmic}
\end{algorithm}
\begin{algorithm}[tbp!]
\caption{Cyclic optimization for known $\sigma^2_n$}
\label{Alg:SLIM_known_sigma}
\begin{algorithmic}[1]
	\REQUIRE  ${\bd}_q^{(0)}$, $\bS$, $\bV$, $\bar{\Omega}_q$, and $\sigma^2_n$
	\ENSURE ${\bd}_{\hat{q}}$
	\STATE Set $n=0$.
	\STATE Set $n=n+1$.
	\STATE Compute $\forall q\in\bar{\Omega}_q$
	\[
	{\bd}_q^{(n)}=\frac{\bP_q^{(n-1)}}{K} \begin{bmatrix} \bv(\theta_1)^\dag \bH\left({\bd}_q^{(n-1)}\right) \bv(\theta_1)
	\\
	\vdots
	\\
	\bv(\theta_L)^\dag \bH\left({\bd}_q^{(n-1)}\right) \bv(\theta_L)
	\end{bmatrix}
	\]
	with $\bH\left({\bd}_q^{(n-1)}\right)$ given by \eqref{eqn:Hdefinition}.
	\STATE Apply Algorithm \ref{Alg:refinement_d}, which returns $\widehat{\bd}_{q}$, $\forall h(q)\in\{1,\ldots,N_{j,\max}\}$, and compute
	\[
	 {\bd}_q^{(n)}=\underset{q\in\bar{\Omega}_q,h(q)\in\{1,\ldots,N_{j,\max}\}}{\arg\dmin}\mbox{BIC}_q(\widehat{\bd}_{q})
	\]
	with BIC$_q$ given by \eqref{eqn:BICq}.
	\STATE If the stopping condition on $n$ is satisfied go to step $6$ else go to step $2$.
	\STATE Return ${\bd}_{\hat{q}}={\bd}_q^{(n)}$.
\end{algorithmic}
\end{algorithm}

Gathering the above estimates, the adaptive LRT can be written as
\be
\Lambda_1(\bZ)=\frac{f_1(\bZ; \tilde{\sigma}_n^2, {\bd}_{\widehat{q}}, H_1)}{f_0(\bZ; \tilde{\sigma}_n^2, \bzero, H_0)}\test\eta,
\label{eqn:adaptiveLRT}
\ee
where ${\bd}_{\widehat{q}} = \widehat{\bd}_{\widehat{q}}$ and $\tilde{\sigma}_n^2$ is an estimate of 
the thermal noise power available at the receiver. For instance, the value of such estimate 
can be an entry of a Lookup Table accounting for different system operating conditions or alternatively, 
it can be periodically computed according to the plan of the system scheduler by disabling the antenna front-end. 
Architecture \eqref{eqn:adaptiveLRT} will be referred to in the following as Sparse Cyclic LRT (SC-LRT).
In the next subsection, we conceive an adaptive procedure which exploits data under test 
to jointly estimate $\sigma_n^2$ and $\bd$ at the price of an additional computational burden.
As a matter of fact, such new procedure comprises two steps which are iterated until a stopping
criterion is satisfied. Specifically, the first step is described in the present subsection, whereas
the second step will be devised in what follows. Thus, the additional computational load is due
to both the second step and the required iterations.

\subsection{Adaptive detector for unknown $\bd$ and $\sigma^2_n$}
\label{Subsec:Unknown_d_sigma}
In this case, both $\sigma_n^2$ and $\bd$ must be estimated from data. While the Maximum Likelihood Estimate (MLE) of $\sigma^2_n$ under $H_0$ can be obtained
in closed-form, the estimation of the unknown parameters under $H_1$ is more problematic and requires elaborate approaches.
To this end, let us note that the estimation procedure for $\bd$ described in the previous subsection, which assumes that $\sigma^2_n$ is known,
can be viewed as a step of a cyclic procedure that, when $\sigma_n^2$ is unknown, 
repeats the following operations
\begin{enumerate}
\item assume that $\sigma_n^2$ is known and estimate $\bd$;
\item assume that $\bd$ is known and estimate $\sigma_n^2$.
\end{enumerate}
Moreover, the estimates obtained at the previous iteration replace the quantities assumed known at the current iteration and so on.
Since the first step of this procedure is described in Subsection \ref{Subsec:knownSigma}, we complete here the procedure by describing
the missing part, namely the estimation of $\sigma_n^2$. 

Thus, let us start assuming that $H_1$ is in force and that an estimate of $\bd$ at the $k$th iteration, $\widehat{\bd}^{(k)}$ say, is available.
Then, compute the MLE of $\sigma_n^2$ for $\bd=\widehat{\bd}^{(k)}$, which is tantamount to solving
\be
\dmax_{\sigma_n^2}\cL(\sigma_n^2),
\ee
where $\cL(\sigma_n^2)=\log f_1(\bZ;\sigma_n^2,\widehat{\bd}^{(k)},H_1)$ is the log-likelihood function for $\bd=\widehat{\bd}^{(k)}$. Now, note that
the $\cL(\sigma_n^2)$ is a continuous function such that
\be
\begin{cases}
\ds\lim_{\sigma^2_n\rightarrow 0^+} \cL(\sigma_n^2) = A < 0,
\\
\ds\lim_{\sigma^2_n\rightarrow +\infty} \cL(\sigma_n^2) = -\infty.
\end{cases}
\ee
As a consequence, the maximum of $\cL(\sigma_n^2)$ occurs at either $\sigma_n=0$ or the local stationary points. In this case, it can be found by
setting to zero the first derivative of $\cL(\sigma_n^2)$ with respect to $\sigma_n^2$, namely
\begin{align}
&\frac{\partial}{\partial \sigma_n^2} \cL(\sigma_n^2) \nonumber
\\
&=\frac{\partial}{\partial \sigma_n^2} \left\{ 
-KN\log\pi - K\log\det\left(\sigma^2_n\bI+\widehat{\bLambda}_d^{(k)}\right)-\tr\left[\left(\sigma^2_n\bI+\widehat{\bLambda}_d^{(k)}\right)^{-1}\bS_d\right]
\right\} \nonumber
\\
&=-K\sum_{i=1}^N \frac{1}{\sigma^2_n+\widehat{\lambda}_{d,i}^{(k)}}+\sum_{i=1}^N 
\frac{\bS_d(i,i)}{\left(\sigma^2_n+\widehat{\lambda}_{d,i}^{(k)}\right)^2}=0 \nonumber
\\
&\Rightarrow \sum_{i=1}^N \frac{\bS_d(i,i)-K\left(\sigma^2_n+\widehat{\lambda}_{d,i}^{(k)}\right)}{\left(\sigma^2_n+\widehat{\lambda}_{d,i}^{(k)}\right)^2}=0,
\label{eqn:derivativeSigma}
\end{align}
where $\widehat{\bLambda}_d^{(k)}\in\R^{N\times N}$ is a diagonal matrix whose nonzero entries 
are the eigenvalues of $\bV\diag\left(\widehat{\bd}^{(k)}\right)\bV^\dag$ denoted by
$\widehat{\lambda}_{d,i}^{(k)}$ with $\widehat{\lambda}_{d,1}^{(k)}\geq \ldots \geq \widehat{\lambda}_{d,N}^{(k)}\geq 0$, 
whereas $\bS_d=\left[\widehat{\bU}_d^{(k)}\right]^\dag \bS \widehat{\bU}_d^{(k)}$ with $\widehat{\bU}_d^{(k)}$ a unitary matrix
whose columns are the eigenvectors of $\bV\diag\left(\widehat{\bd}^{(k)}\right)\bV^\dag$ corresponding to $\widehat{\lambda}_{d,i}^{(k)}$, $i=1,\ldots,N$. 
Now, by {\em Abel-Ruffini Theorem} \cite{pesic2004abel}, when $N\geq 3$, equation \eqref{eqn:derivativeSigma} does not admit solutions 
in algebraic form. For this reason, we solve it resorting to numerical
routines and choose the positive solution, $\left(\widehat{\sigma}_{n,1}^2\right)^{(n+1)}$ say, 
greater than or equal to $1$ and that returns the highest value of $\cL(\sigma_n^2)$.
If such solution does not exist, we set $\left(\widehat{\sigma}_{n,1}^2\right)^{(n+1)}=1$.
Once $\left(\widehat{\sigma}_{n,1}^2\right)^{(n+1)}$ is available, we exploit the cyclic optimization of Subsection \ref{Subsec:knownSigma} to compute
$\widehat{\bd}^{(k+1)}$ where $\sigma_n^2$ is replaced by $\left(\widehat{\sigma}_{n,1}^2\right)^{(k+1)}$. 
The entire procedure, summarized in Algorithm \ref{Alg:SLIM_unknown_sigma}, can terminate after a fixed number of iterations or 
when a convergence criterion is satisfied as, for instance,
\be
\frac{\| \widehat{\bd}^{(k)} - \widehat{\bd}^{(k-1)} \|}{\| \widehat{\bd}^{(k-1)} \|}+
\frac{| \left(\widehat{\sigma}_{n,1}^2\right)^{(k)} - \left(\widehat{\sigma}_{n,1}^2\right)^{(k-1)}|}
{ \left(\widehat{\sigma}_{n,1}^2\right)^{(k-1)} }< \epsilon,
\label{eq_stopcriterion}
\ee
with $\epsilon$ a suitable small positive number.

On the other hand, under $H_0$, it is not difficult to show that the MLE of $\sigma_n^2$ is given by
\be
\widehat{\sigma}_{n,0}^2 = \frac{1}{KN}\tr[\bS]
\ee
and replacing the above estimates in the LRT, we come up with
\be
\Lambda_2(\bZ)=\frac{f_1(\bZ; \widehat{\sigma}_{n,1}^2, \widehat{\bd}, H_1)}{f_0(\bZ; \widehat{\sigma}_{n,0}^2, \bzero, H_0)}\test\eta,
\ee
where $\widehat{\sigma}_{n,1}^2$ and $\widehat{\bd}$ are the final estimates provided by Algorithm \ref{Alg:SLIM_unknown_sigma}. In what follows, we refer to the above decision scheme as Sparse 
Doubly Cyclic LRT (SDC-LRT).

\begin{algorithm}[tbp!]
\caption{Cyclic optimization for unknown $\sigma_n^2$}
\label{Alg:SLIM_unknown_sigma}
\begin{algorithmic}[1]
	\REQUIRE  ${\bd}_q^{(0)}$, $\bS$, $\bV$, $\bar{\Omega}_q$, and $(\sigma^2_{n,1})^{(0)}$ .
	\ENSURE $\widehat{\bd}$ and $\widehat{\sigma}^2_{n,1}$.
	\STATE Set $n=0$.
	\STATE Set $n=n+1$.
	\STATE Execute steps $3$ and $4$ of Algorithm \ref{Alg:SLIM_known_sigma} setting $\sigma^2_n=(\sigma^2_{n,1})^{(n-1)}$ to obtain ${\bd}_q^{(n)}$.
	\STATE Compute the eigendecomposition of $\bV\diag({\bd}_q^{(n)})\bV^\dag$.
	\STATE Compute $(\sigma^2_{n,1})^{(n)}$ as the solution of
	\[
	 \sum_{i=1}^N \frac{\bS_d(i,i)-K\left(\sigma^2_n+\widehat{\lambda}_{d,i}^{(k)}\right)}{\left(\sigma^2_n+\widehat{\lambda}_{d,i}^{(k)}\right)^2}=0
	\]
	that maximizes $\cL(\sigma_n^2)$.
	\STATE If the stopping condition on $n$ is satisfied go to step $7$ else go to step $2$.
	\STATE Return $\widehat{\bd}={\bd}_q^{(n)}$ and $\widehat{\sigma}^2_{n,1}=(\sigma^2_{n,1})^{(n)}$.
\end{algorithmic}
\end{algorithm}

\bigskip

Before concluding this section and presenting the numerical examples, we highlight that the estimate 
of $\bd$, $\widehat{\bd}$ say, can be used to infer the number of NLJs and their AOAs. However, $\widehat{\bd}$
may contain false objects (ghosts) induced by the energy spillover. In order to mitigate the number of ghosts,
we apply an additional thresholding of the entries of $\widehat{\bd}$ and we resort to 
the same fusion strategy 
proposed in \cite{ECCMYan}, where the grid used to sample the angular 
sector under surveillance is partitioned into subsets associated to a specific AOA and the entries 
of $\widehat{\bd}$ falling in a subset are merged together. The interested reader is referred to \cite{ECCMYan}
for further details. Finally, it is clear that other fusion strategies are possible leading to
better estimation and/or classification performance especially in the case where the actual AOAs of the NLJs
are in between the points of the sampling grid. 
For instance, an interpolation of consecutive nonzero entries of $\widehat{\bd}$ 
can be performed, whereupon the resulting peaks can be selected. Another approach would consist in
increasing the angular resolution of the grid in the sectors that contain consecutive nonzero entries 
of $\widehat{\bd}$. As a result, the actual AOAs of the NLJs are very close to the oversampled grid points.
The design of different fusion strategies is out of the scope of the present paper and represents the current
research line.

\section{Illustrative Examples and Discussion}
\label{Sec:Performance}
In this section, we present some numerical examples aimed at showing the detection and estimation
capabilities of the SDC-LRT and the SC-LRT for known\footnote{Comparing SC-LRT for known $\sigma^2_n$
with the SDC-LRT allows us to quantify the loss due to the estimation of $\sigma^2_n$.} $\sigma^2_n$. 
For comparison purpose, we also assess the performance of the LRT where the unknown parameters are
estimated by means of the SParse Iterative Covariance-based Estimation (SPICE) algorithm
whose theoretical formulation is laid down in \cite{StoicaSPICE} and that is well-suited to 
the covariance matrix model at hand given by \eqref{eqn:covarianceModel}. This competitor
will be denoted by the acronym SPICE-LRT.
Two operating scenarios are considered, which differ
in the number of NLJs. Specifically, the former contains $N_j=3$ NLJs, whereas the latter is characterized
by the presence of $N_j=4$ NLJs. In both scenarios, NLJs share the same (nominal) power and 
are located within an angular
sector under surveillance ranging from $-22^{\circ}$ to $22^{\circ}$ and uniformly sampled at $1$ degree,  
$2$ degrees, or $3$ degrees. 
The nominal AOA of the NLJs, measured with respect to the array normal, are assumed to lie on the
sampling grid (``on-grid'' case) and are given by
\begin{itemize}
\item $\bar{\theta}_{1}=-10^{\circ}, \bar{\theta}_{2}=-4^{\circ}$ and $\bar{\theta}_{3}=8^{\circ}$ for a 
spatial sampling rate of $1$ degree, $2$ degree, and $3$ degree in the scenario which assumes $N_j=3$ NLJs;
\item $\bar{\theta}_{1}=-10^{\circ}, \bar{\theta}_{2}=-4^{\circ}$, $\bar{\theta}_{3}=8^{\circ}$, 
and $\bar{\theta}_{4}=14^{\circ}$ for a spatial sampling rate of $1$ degree, $2$ degree, and $3$ degree 
in the other scenario which assumes $N_j=4$ NLJs.
\end{itemize}
Besides, we also consider the ``off-grid'' situation where the actual AOAs of the NLJs are in between
the grid samples (a point better explained below).
Finally, we show that at high NLJ power, the proposed algorithms provide high-quality estimates 
of the NLJ parameters.

The Jammer-to-Noise Ratio (JNR) is defined as
$\mbox{JNR}=\frac{\sigma_j^2}{\sigma^2_n}$
with $\sigma^2_n=2$. The analysis relies on the following figures of merit:
\begin{itemize}
\item the Probability of Jammer Detection ($P_{jd}$) for a given $P_{fjd}$;
\item the Root Mean Square (RMS) value for the number of missed NLJs, the number of ghost NLJs
and the Hausdorff metric \cite{4567674} between\footnote{Note that such figures of merit make 
sense when the performance are evaluated on-grid assumption. Conversely, in the case
of off-grid angular positions, another figure of merit must be considered.} $\bd$ and $\bd_{\hat{q}}$. 
The latter belongs to the family of the multi-object distances which are able to capture the error 
between two sets of vectors and is defined 
as $h_d(\cX,\cY) = \max \{ \max_{x \in \cX} \min_{y\in \cY} d(x,y) , \max_{y\in \cY} \min_{x\in \cX} d(x,y)  \}$ 
with $\cX$ and $\cY$ are the sets of the coordinates of the nonzero entries of $\bd$ and $\bd_{\hat{q}}$, 
respectively (these figures of merit are computed exploiting the fusion strategy of \cite{ECCMYan} 
with a subset cardinality equal to $3$);
\item the classification histograms (computed exploiting the fusion strategy of \cite{ECCMYan} with a 
subset cardinality equal to $3$) namely the percentages of declaring that $n$, $n=1,\ldots,6$,
NLJs are present when the actual number of NLJs is either $3$ or $4$;
\item the RMS values for the angular error between the actual AOA and the estimated direction closest
to the former (off-grid case only).
\end{itemize}
Due to the lack of closed-form expressions for the above metrics, we resort to standard Monte Carlo counting
techniques. Specifically, the detection thresholds are computed over $100/P_{fjd}$ independent trials with
$P_{fjd}=10^{-2}$, whereas the classification percentages and the RMS values
are estimated exploiting $1000$ independent trials.
In the case of off-grid NLJ angular positions, at each Monte Carlo trial, 
the AOAs are generated as independent random variables uniformly distributed in 
$[\bar{\theta}_i-1, \ \bar{\theta}_i+1]$ or $[\bar{\theta}_i-\Delta\theta/2, \ \bar{\theta}_i+\Delta\theta/2]$ 
degrees, $i=1,\ldots,3$ or $i=1,\ldots,4$, where $\Delta\theta$ is the grid sampling interval.
It is worth noticing that this off-grid analysis is aimed at illustrating the behavior of the newly proposed method
in three different situations, namely an unfavorable case where the grid is sampled at $1$ degree 
(and, hence, the variation range of the actual direction for each jammer comprises three grid points), 
a favorable case where the grid is sampled at $3$ degrees (i.e., the actual direction of each jammer is very 
close to a nominal grid point), and an intermediate situation with a sampling interval of $2$ degrees. 
From an operating point of view, it would be possible to bring back to one of the above cases by oversampling
the angular sectors identified by a preliminary application of the estimation procedure over a rough search grid.
Moreover, as already stated, we perform an additional thresholding of the entries of $\bd_{\hat{q}}$.
To this end, the threshold is set to ensure a probability of declaring the presence of spurious NLJs
equal to $10^{-3}$. Finally, all the numerical examples assume $N=32$ and $K=64$, whereas
the estimation procedures terminate when the convergence criterion in \eqref{eq_stopcriterion} 
is satisfied with $\epsilon=10^{-2}$.

\subsection{First Operating Scenario: $3$ NLJs}
\begin{figure}[tbp!]
    \centering
    \includegraphics[width=.7\textwidth]{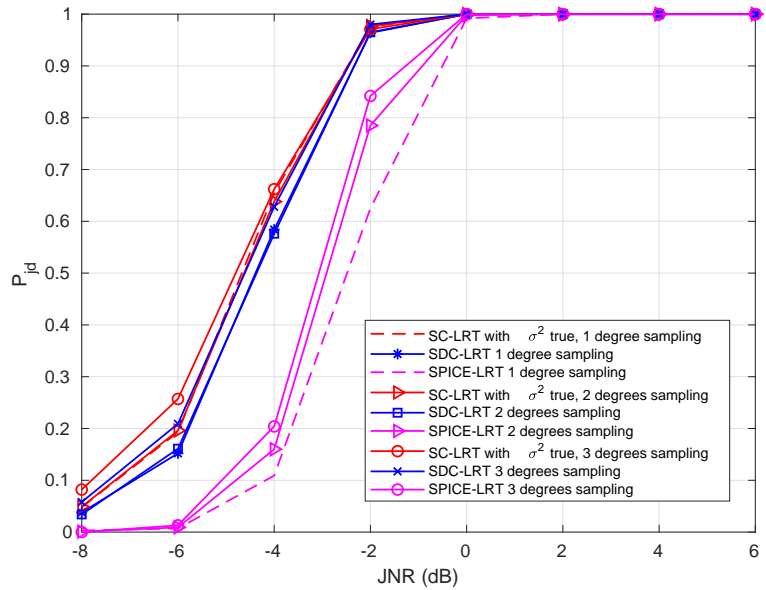}
    \caption{$P_{jd}$ versus JNR for the SC-LRT, the SDC-LRT, and the SPICE-LRT assuming $N_j=3$ and the 
    nominal AOAs for the NLJs.}
    \label{p1}
\end{figure}

Let us start the analysis by focusing on the scenario that contains $N_j=3$ NLJs. 
In Figure \ref{p1}, we plot the $P_{jd}$ of the decision schemes devised in Subsections
\ref{Subsec:knownSigma} and \ref{Subsec:Unknown_d_sigma} along with that of the SPICE-LRT
assuming that the AOAs of the NLJs belong to the sampling grid.
Inspection of the figure highlights that the performance improves as the sampling interval grows. 
This behavior can be motivated by noticing that a wider sampling interval would decrease 
the coherence of the dictionary $\bV$ leading, as a consequence, to an improvement of the estimation 
quality of $\bd$ \cite{CompressiveSampling}.
Moreover, the proposed detectors exhibit a gain of about $2$ dB at $P_{jd}=0.9$ with respect to the SPICE-LRT.
It is also worth observing that the $P_{jd}$ for SC-LRT and SDC-LRT 
achieves values greater than $0.9$ for $\mbox{JNR}$ values greater
than about $-2$ dB.

\begin{figure}[tbp!]
    \centering
    \includegraphics[width=1\textwidth]{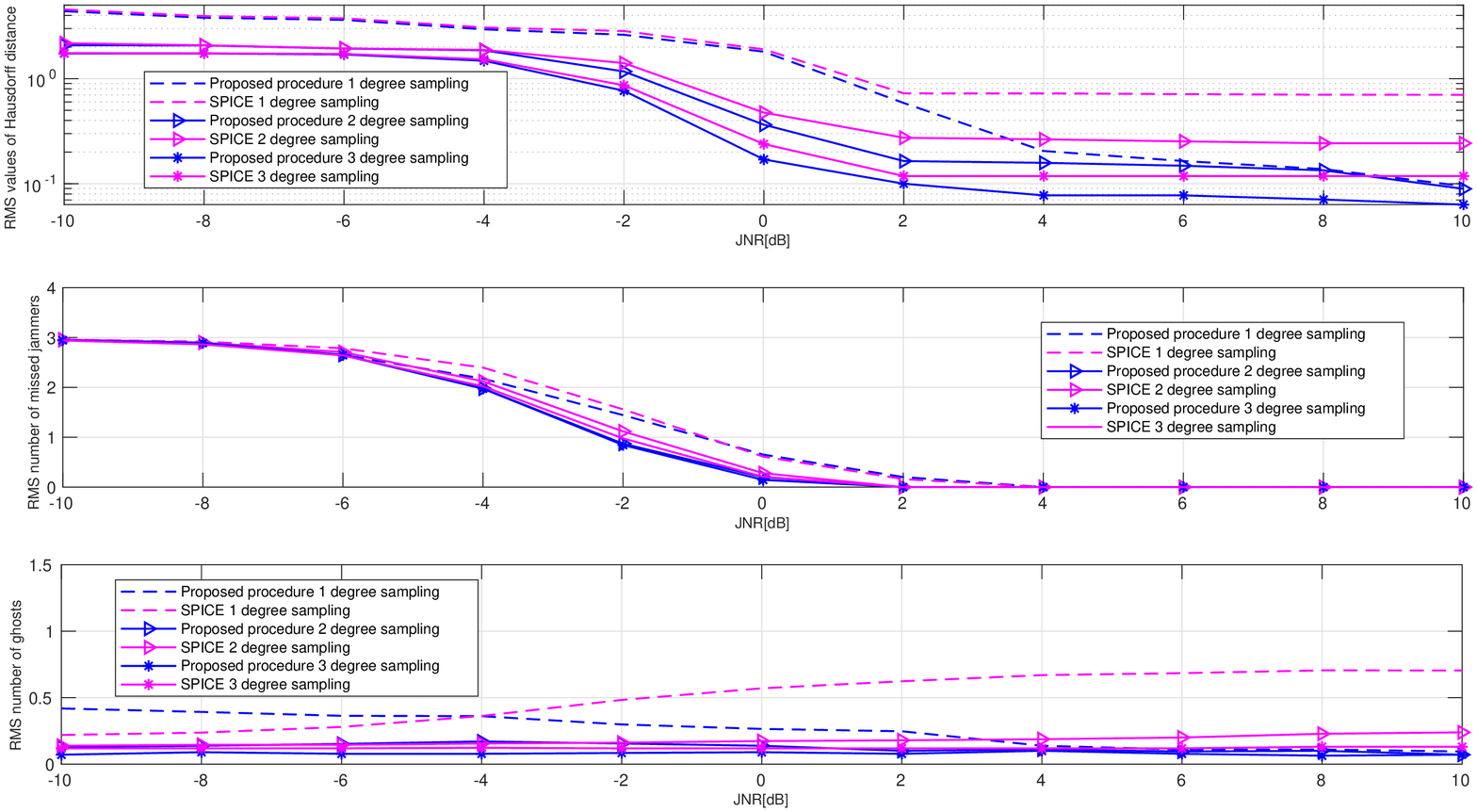}
    \caption{RMS value for the Hausdorff distance, number of missed jammers, and number of ghosts versus 
    JNR assuming $N_j=3$ and the nominal AOAs for the NLJs.}
    \label{p2}
\end{figure}

\begin{figure}[tbp!]
    \centering
    \includegraphics[width=.6\textwidth]{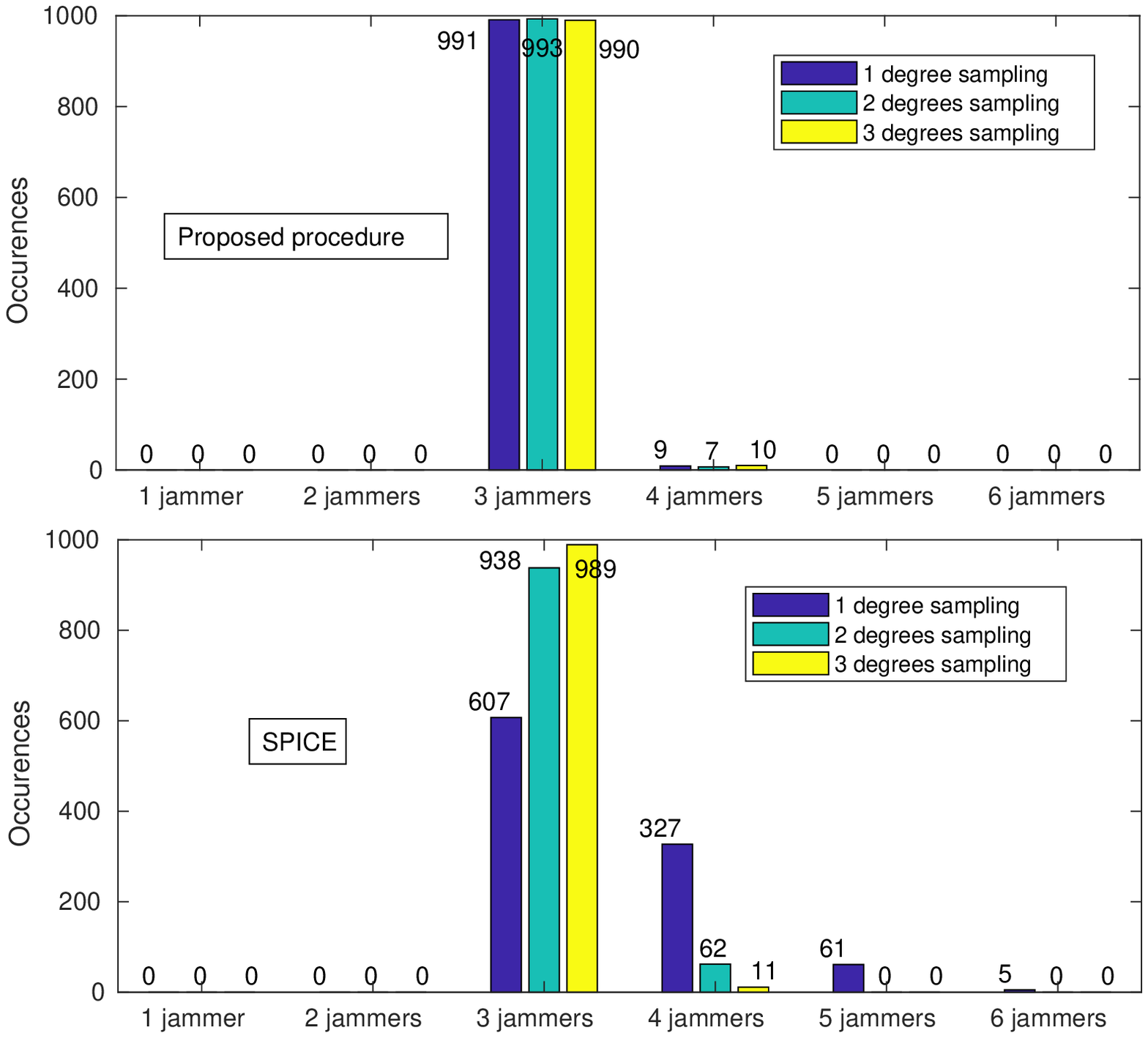}
    \caption{{Classification histograms for the number of times that the procedures return $1$ jammer,$\ldots$, $6$
    jammers assuming $\mbox{JNR}=10$ dB, $N_j=3$, and the nominal AOAs for the NLJs.}}
    \label{p3}
\end{figure}

In Figure \ref{p2}, the RMS values for the Hausdorff distance between the true and estimated $\bd$, 
the number of missed jammers, and
the number of ghosts against the JNR are plotted for the same parameter values as in Figure \ref{p1}. 
Both the SPICE and the proposed procedure (Algorithm \ref{Alg:SLIM_unknown_sigma}) 
exhibit excellent performance rates which improve as
the JNR increases. However, to be more precise, the proposed procedure performs slightly better than SPICE
for the considered figures of merit and the parameter setting. As a matter of fact, for high JNR values
the Hausdorff distance provided by SPICE is biased whereas that related to the proposed procedure
is strictly decreasing as the JNR increases. The differences observed in the last two subfigures are less
evident except for the RMS number of ghosts when the grid is sampled at $1$ degree.
In order to provide a complete picture of the performance for the on-grid case comprising $3$ NLJs, 
Figure \ref{p3} shows the classification 
histograms assuming $\mbox{JNR}=10$ dB (recall that NLJs transmit very high power) and the nominal AOA for 
the NLJs. More precisely, such histograms count the number of times that the estimation procedures state 
that $n\in\{1,\ldots,6\}$ NLJs are present (recall that the ground-truth is $3$ NLJs). 
It turns out that the proposed procedure can guarantee a percentage
of correct estimation for the number of NLJs greater than $99$ \% for all the considered sampling intervals, 
whereas SPICE exhibits percentage higher than
$93$ \% when the grid sampling rate is $2$ or $3$ degrees. For the case of a grid sampled at $1$ degree,
the percentage of correct estimation for SPICE decreases to about $60$ \% yielding a nonnegligible 
overestimation attitude.

\begin{figure}[tbp!]
    \centering
    \includegraphics[width=.7\textwidth]{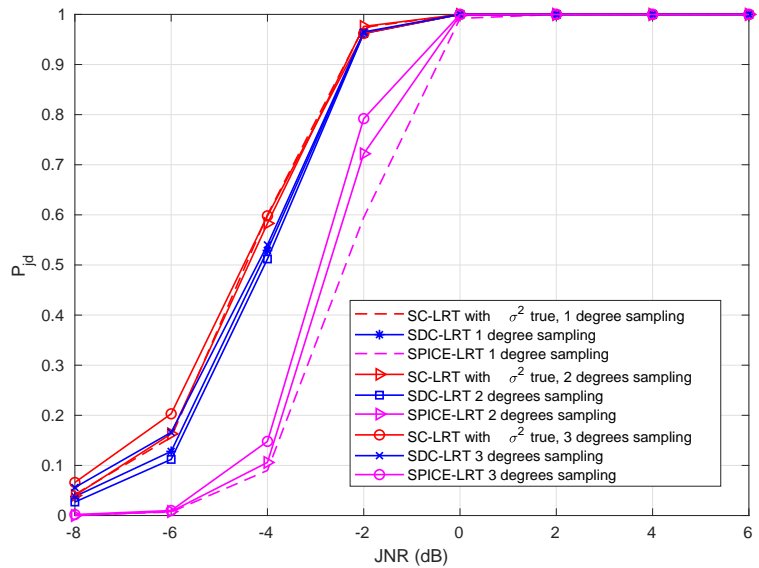}
    \caption{$P_{jd}$ versus JNR for the SC-LRT, the SDC-LRT, and the SPICE-LRT assuming $N_j=3$, the 
    nominal AOAs for the NLJs, and a JNR variation of $5$ dB during data acquisition.}
    \label{fig:Pjd_power}
\end{figure}

The next illustrative example assumes that the NLJs have not yet transmitted their maximum power level when
the radar is forming the set of data under test. Specifically, for each nominal JNR value, we generate
$\bz_1$ with a JNR given by the nominal value minus $5$ dB, then the remaining $\bz_i$s, $i>1$, are 
built up increasing the initial JNR by $1$ dB until the nominal value is achieved. 
Figure \ref{fig:Pjd_power}, where the $P_{jd}$ is shown as a function of the JNR
for this scenario, confirms the ranking observed in Figure \ref{p1} with a slight performance degradation, 
which is, nevertheless, expected since the actual amount of collected energy is less than the nominal value. 
As for the other figures of merit, results not reported here for brevity are aligned with what 
observed in Figures \ref{p2} and \ref{p3}.

\begin{figure}[tbp!]
    \centering
    \includegraphics[width=.7\textwidth]{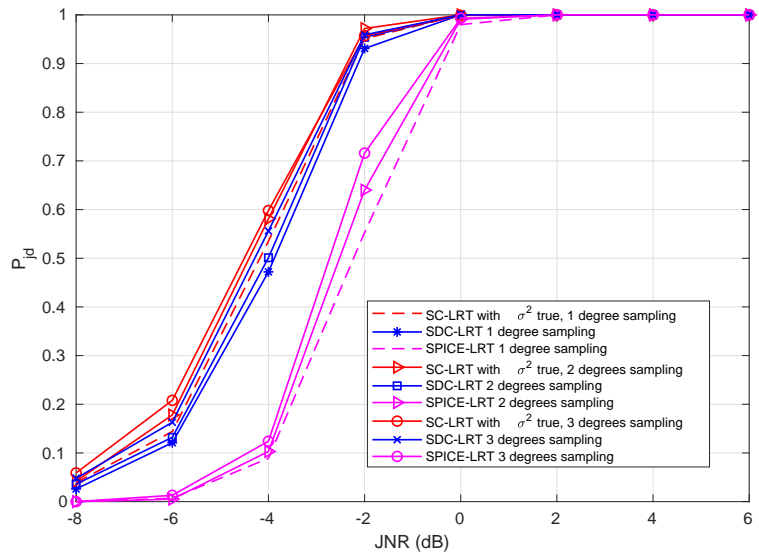}
    \caption{{$P_{jd}$ versus JNR for the SC-LRT, the SDC-LRT, and the SPICE-LRT assuming $N_j=3$ and the 
    AOAs of the NLJs in between the sampling grid points.}}
    \label{fig:Pjd_offgrid}
\end{figure}

\begin{figure}[tbp!]
    \centering
    \includegraphics[width=.7\textwidth]{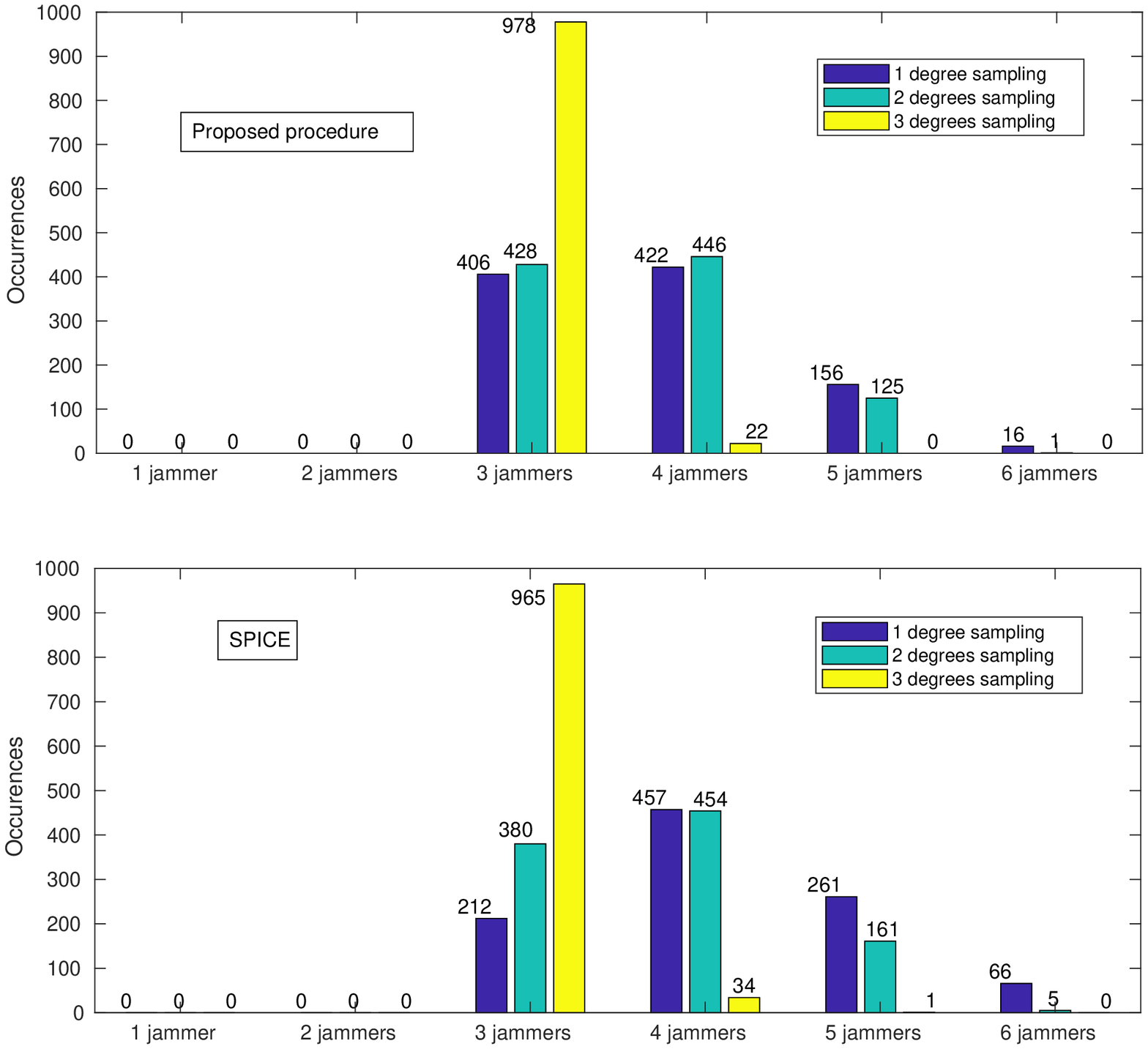}
    \caption{Classification histograms for the number of times that the procedures return $1$ jammer,$\ldots$, $6$
    jammers assuming $\mbox{JNR}=10$ dB, $N_j=3$, and the AOAs of the NLJs in between the sampling grid points.}
    \label{fig:hist_offgrid}
\end{figure}

\begin{figure}[tbp!]
    \centering
    \includegraphics[width=.8\textwidth]{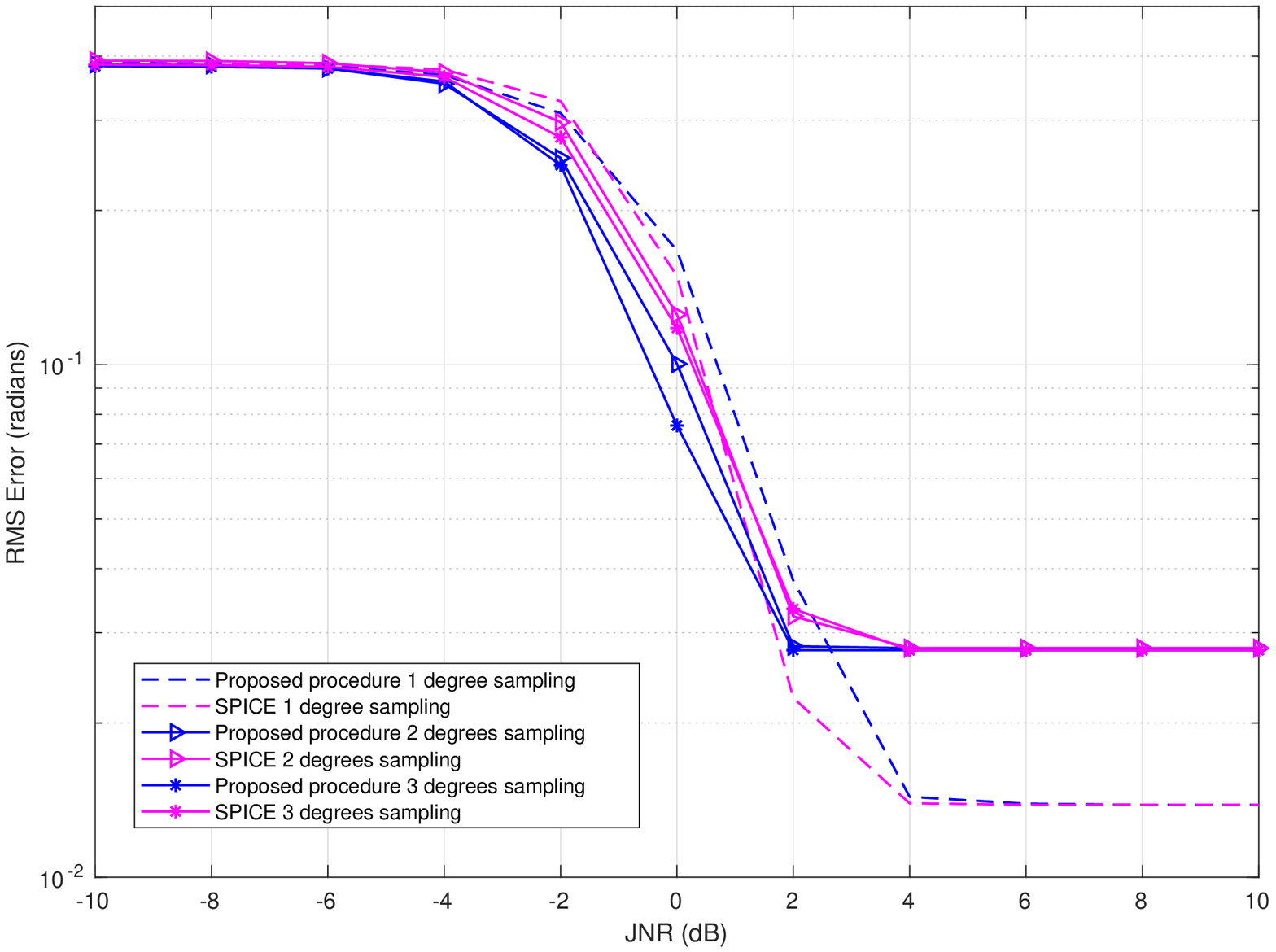}
    \caption{RMS error between the actual AOA of the NLJs and the estimated direction closest to the former 
    versus the JNR assuming $N_j=3$ and the AOAs of the NLJs in between the sampling grid points.}
    \label{fig:RMSE_offgrid}
\end{figure}

\begin{figure}[tbp!]
    \centering
    \includegraphics[width=.8\textwidth]{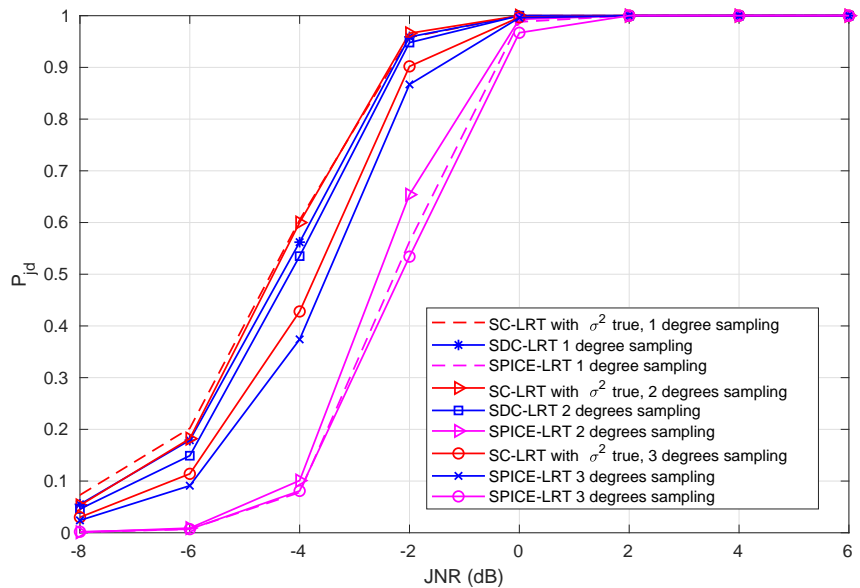}
    \caption{{$P_{jd}$ versus JNR for the SC-LRT, the SDC-LRT, and the SPICE-LRT assuming $N_j=3$ and the 
    AOAs of the NLJs uniformly generated in a window of size the sampling interval.}}
    \label{fig:Pjd_offgrid_halfsize}
\end{figure}

\begin{figure}[tbp!]
    \centering
    \includegraphics[width=.8\textwidth]{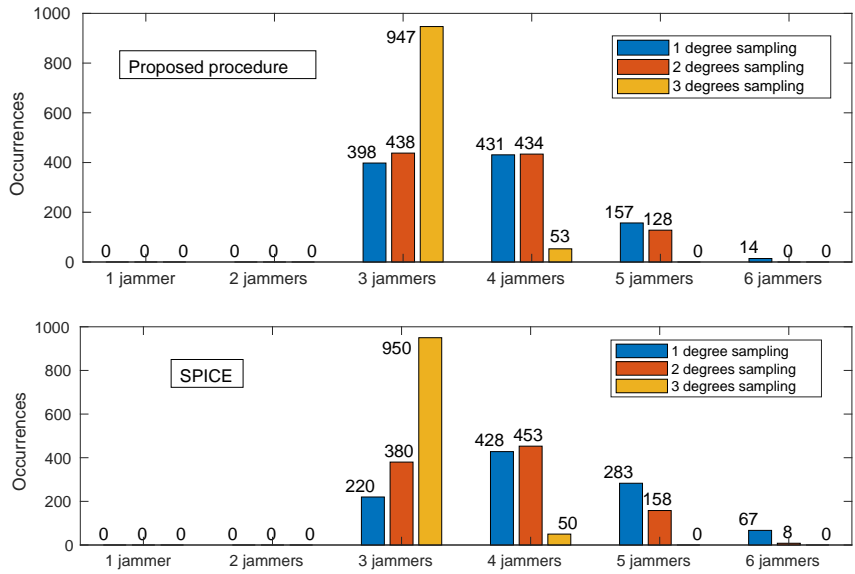}
    \caption{Classification histograms for the number of times that the procedures return $1$ jammer,$\ldots$, $6$
    jammers assuming $\mbox{JNR}=10$ dB, $N_j=3$, and the AOAs of the NLJs uniformly generated in a 
    window of size the sampling interval.}
    \label{fig:hist_offgrid_halfsize}
\end{figure}

Now, we focus on the case where the actual angular positions of the three NLJs are in between the grid points.
In this case, besides the detection performance, we consider the classification histograms and RMS value of the
angular difference between the actual position of the NLJs and estimated direction which is closest
to the former. Note that the other figures of merit do not make sense in this case. 
The detection performance is shown in Figure \ref{fig:Pjd_offgrid} where an overall loss of at most $0.5$ dB 
at $P_{jd}=0.9$ and for the curves related to the sampling intervals of $2$ and $3$ degrees can 
be measured with respect to Figure \ref{p1}. On the other hand, the curves representative of the 
grid sampled at $1$ degree remain unaltered. The figure also confirms that the
SC-LRT and SDC-LRT are superior to the SPICE-LRT with a gain of about $2$ dB and, in addition, that
a wider sampling interval leads to slightly better performance. The classification histograms under the
off-grid assumption are shown in Figure \ref{fig:hist_offgrid}, where, as expected, a sampling interval
of $3$ degrees enhances the estimation quality for the number of NLJs since it decreases the energy spillover
of the NLJs between consecutive grid points. In this case, the proposed procedure is slightly superior to
the SPICE algorithm. For a sampling interval of $1$ and $2$ degrees, both algorithms provide 
a percentage of correct classification less than $50$ \% with the SPICE algorithm being more inclined to
overestimate the number of NLJs than the proposed procedure. 
For instance, note that for a sampling interval of $1$ degree, the sum of the occurrences for the SPICE is less
than $1000$, because the latter in some Monte Carlo trials returns a number of NLJs greater than $6$.
In the Figure \ref{fig:RMSE_offgrid},
we plot the RMS angular distance between the actual AOA and the estimated AOA closest to the former versus
the JNR. The figure points out that the considered procedures share the same performance and, more precisely,
for JNR values greater than $2$ dB the RMS error is less than $2$ degrees. Importantly, a grid sampled at $3$ degrees
allows for RMS values less than $1$ degree for JNR$\geq 4$ dB.
Finally, in Figures \ref{fig:Pjd_offgrid_halfsize}-\ref{fig:hist_offgrid_halfsize}, 
we compute the detection curves 
and the classification histograms by assuming that the actual NLJ angular positions are uniformly 
distributed in a window centered on the nominal AoA and of size exactly equal to the sampling 
interval. The behavior observed in these last figures is aligned with that described before confirming
the superiority of the proposed method over SPICE.

\begin{figure}[tbp!]
    \centering
    \includegraphics[width=.9\textwidth]{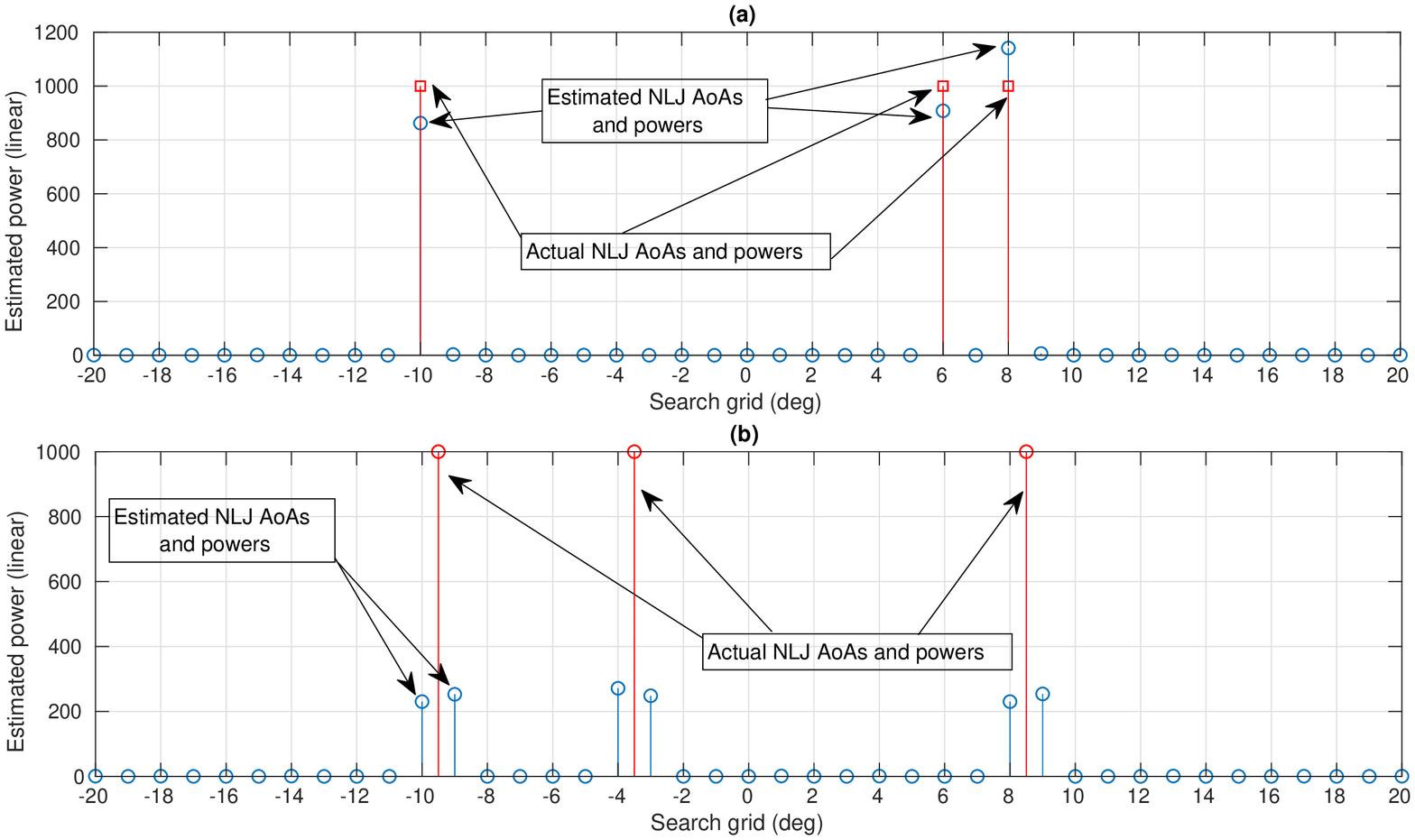}
    \caption{Estimated power (single snapshot) versus search grid angles for three jammers sharing JNR$=30$ dB
    located at: $-10^\circ$, $6^\circ$, and $8^\circ$ subplot (a); $-9.5^\circ$, $-3.5^\circ$, 
    and $8.5^\circ$ subplot (b).}
    \label{fig:singleSnapshot}
\end{figure}

The last illustrative example (Figure \ref{fig:singleSnapshot}) 
of this subsection shows that the estimation quality is high in the case of large NLJ 
power values. To this end, we show the returned estimates for two outcomes of two Monte Carlo trials.
Specifically, in Subplot (a), we plot the interference power estimates as a function of the angles 
belonging to the search 
grid sampled at $1$ degree for three jammers at $\bar{\theta}_1=-10$, $\bar{\theta}_2=6$, and $\bar{\theta}_3=8$ 
with JNR=$30$ dB; Subplot (b) shares the same parameter setting as Subplot (a) but for the NLJ AoAs,
which are $\bar{\theta}_1=-9.5$, $\bar{\theta}_2=-3.5$, and $\bar{\theta}_3=8.5$. Inspection of Subplot (a)
highlights the enhanced resolution provided by the sparse approach along with a significant attenuation
of ``sidelobe'' effects, whereas in Subplot (b), the spillover of the NLJ power between adjacent grid points 
can be observed motivating the need of suitable fusion strategies.


\subsection{Second Operating Scenario: $4$ NLJs}
In this last subsection, we repeat previous analysis assuming that $N_j=4$ NLJs are present.
This analysis is aimed at investigating the effect of an increase of the NLJ number on the performance.

\begin{figure}[tbp!]
    \centering
    \includegraphics[width=.8\textwidth]{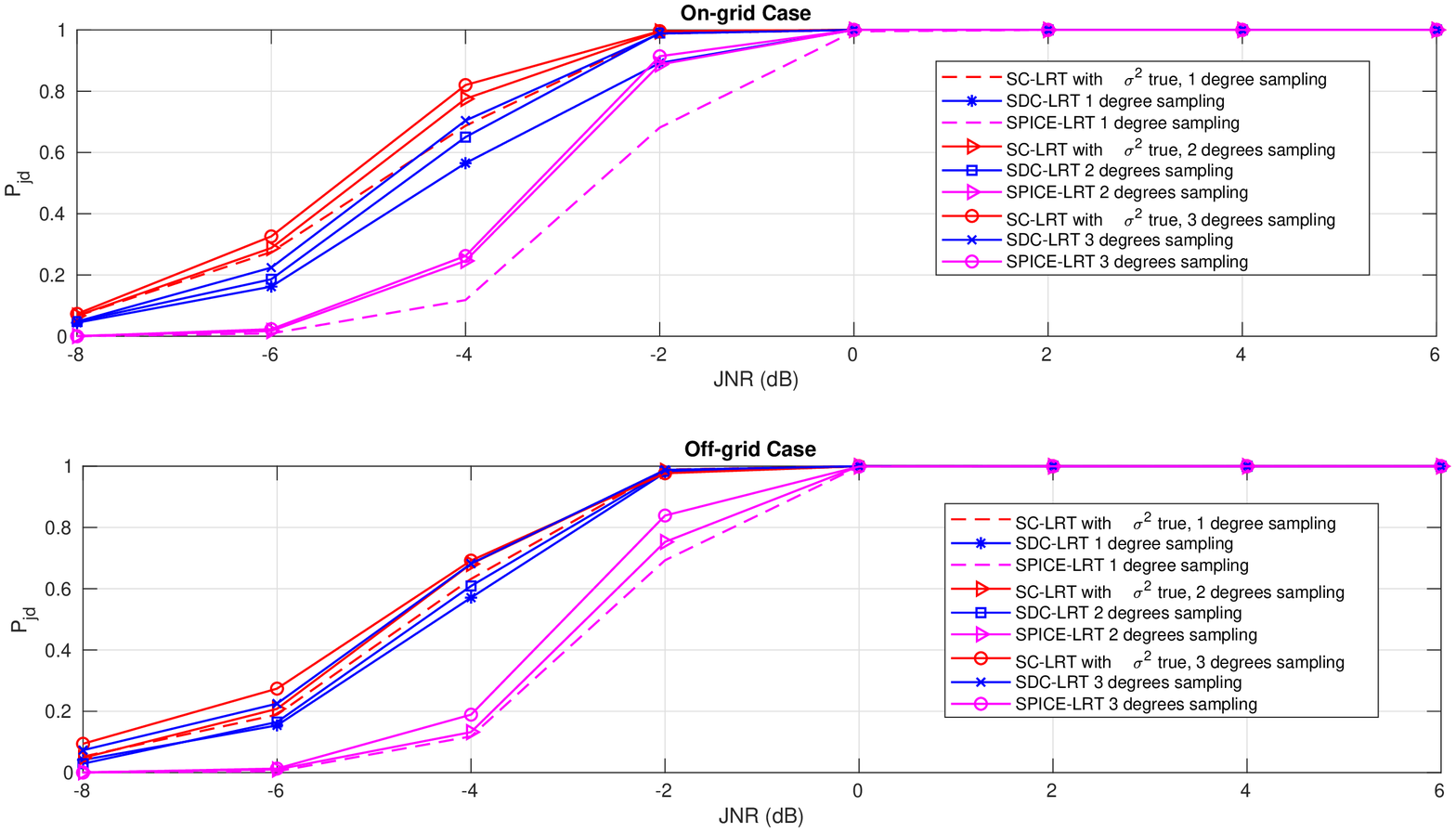}
    \caption{$P_{jd}$ versus JNR for the SC-LRT, the SDC-LRT, and the SPICE-LRT assuming $N_j=4$.}
    \label{fig:case_2_Pjd}
\end{figure}

\begin{figure}[tbp!]
    \centering
    \includegraphics[width=1\textwidth]{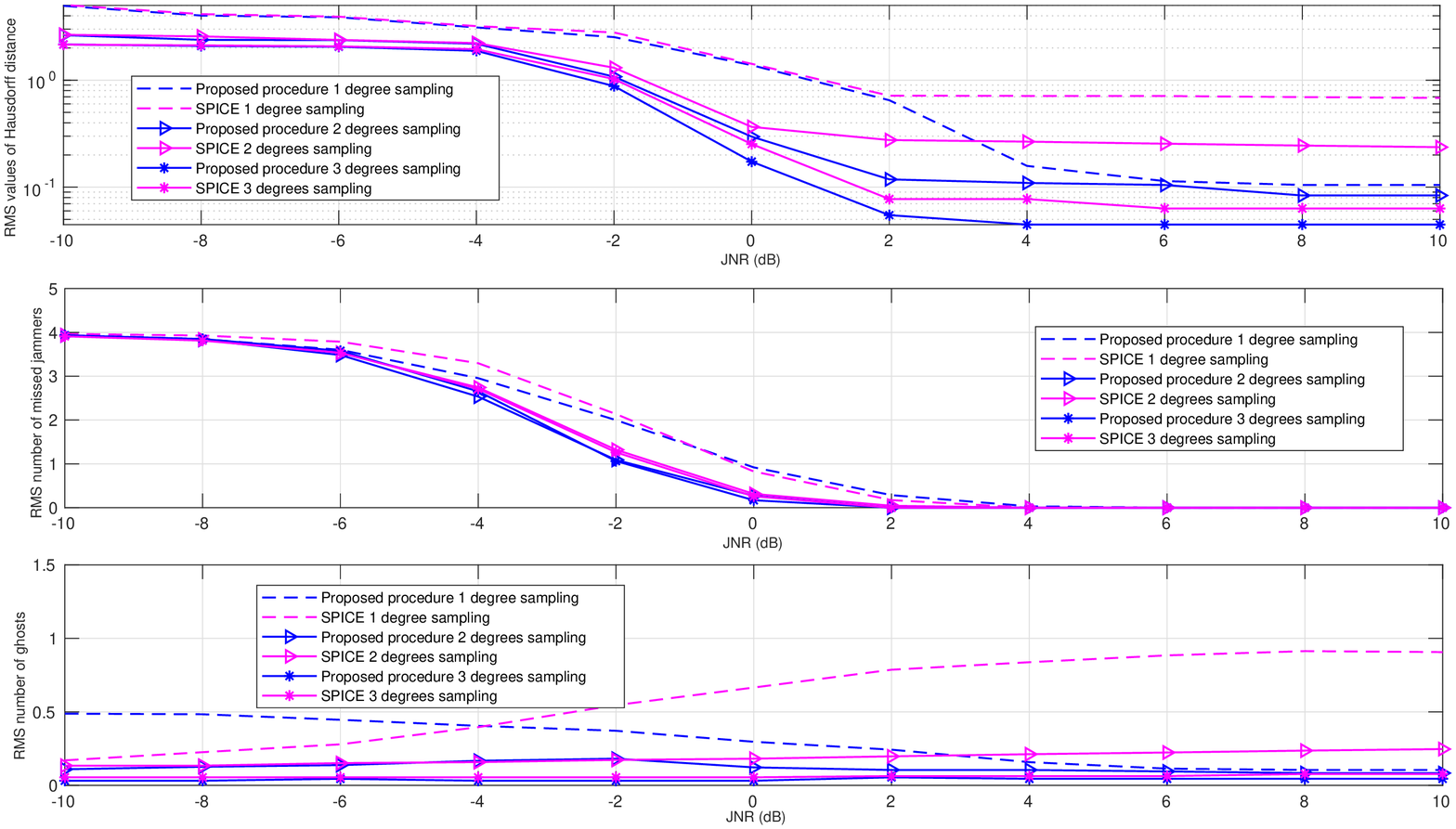}
    \caption{RMS value for the Hausdorff distance, number of missed jammers, and number of ghosts versus 
    JNR assuming $N_j=4$ and the nominal AOAs for the NLJs.}
    \label{fig:case_2_EstimationPerf_ongrid}
\end{figure}

\begin{figure}[tbp!]
    \centering
    \includegraphics[width=.8\textwidth]{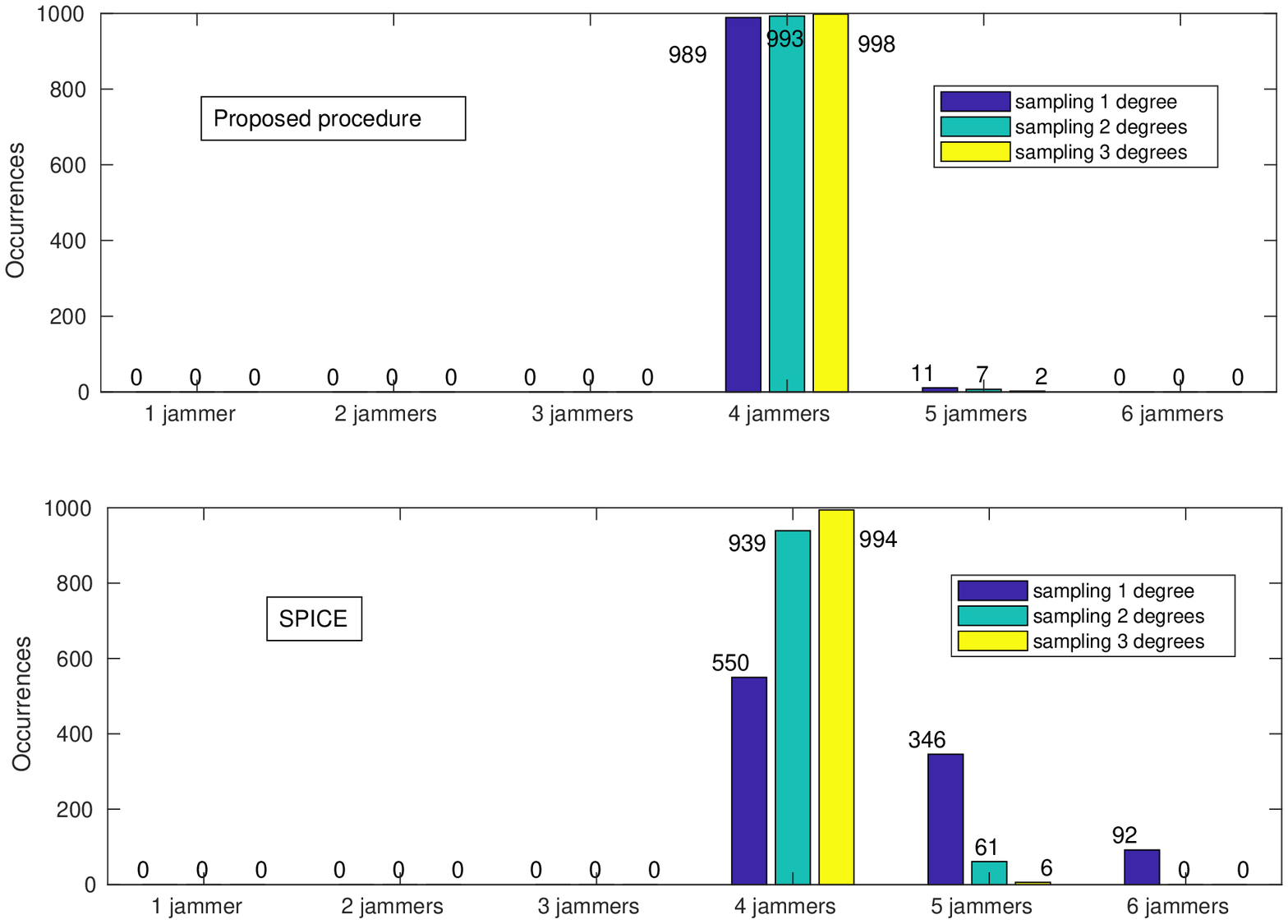}
    \caption{Classification histograms for the number of times that the procedures return $1$ jammer,$\ldots$, $6$
    jammers assuming $\mbox{JNR}=10$ dB, $N_j=4$, and the nominal AOAs for the NLJs.}
    \label{fig:case2_hist}
\end{figure}

\begin{figure}[tbp!]
    \centering
    \includegraphics[width=1\textwidth]{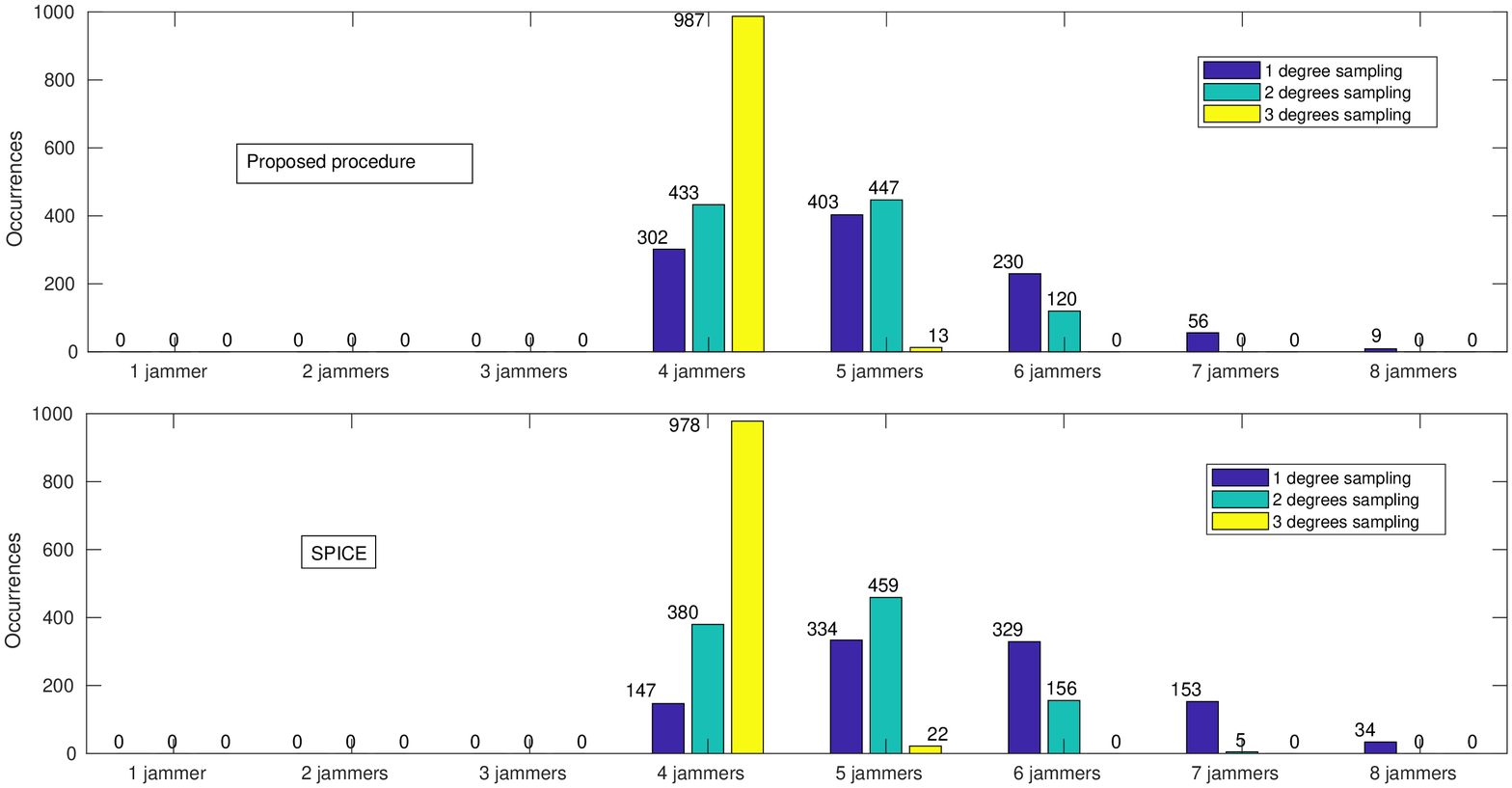}
    \caption{Classification histograms for the number of times that the procedures return $1$ jammer,$\ldots$,$8$
    jammers assuming $\mbox{JNR}=10$ dB, $N_j=4$, and the AOAs of the NLJs in between the sampling grid points.}
    \label{fig:case2_hist_offgrid}
\end{figure}

\begin{figure}[tbp!]
    \centering
    \includegraphics[width=.8\textwidth]{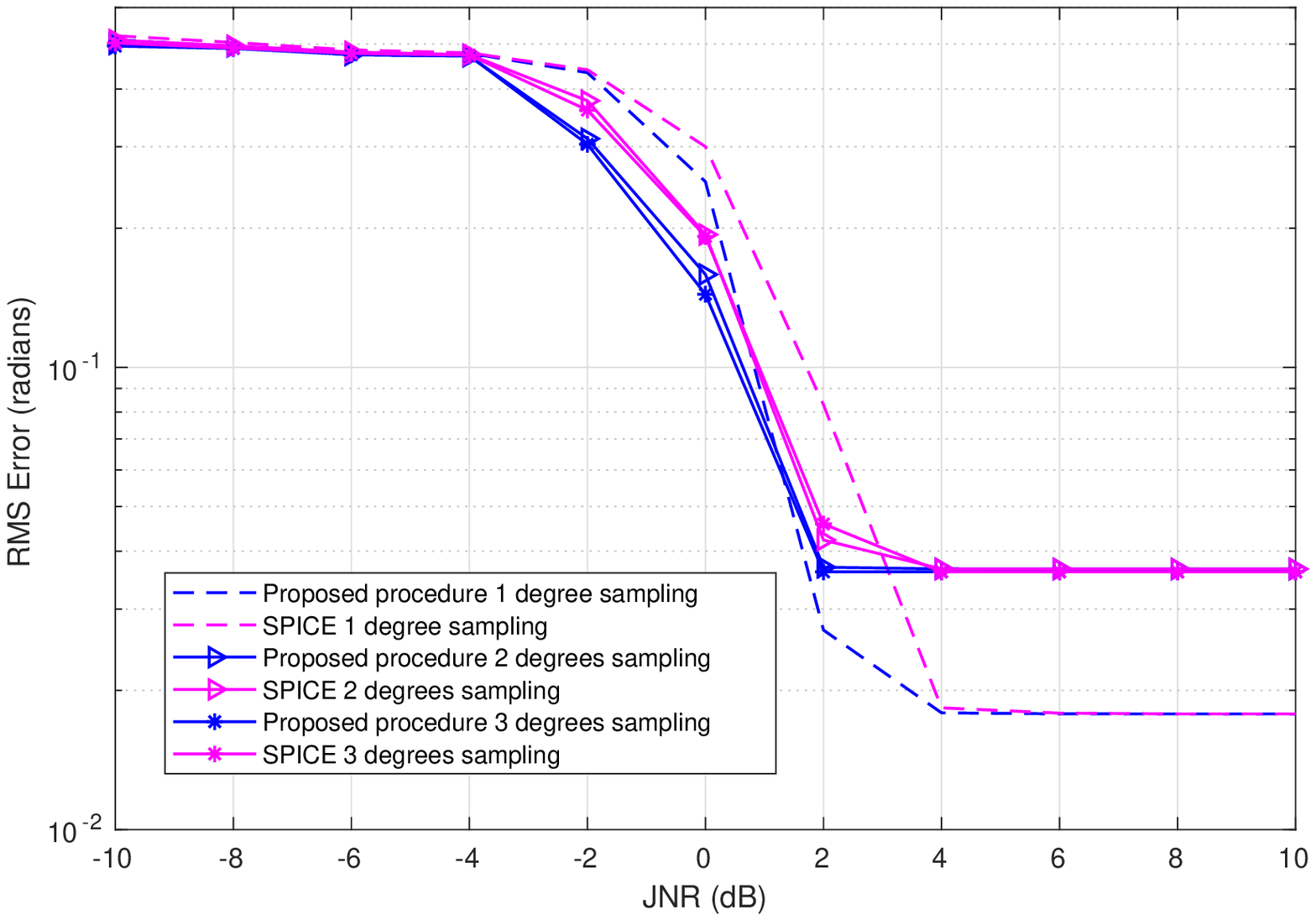}
    \caption{RMS error between the actual AOA of the NLJs and the estimated direction closest to the former 
    versus the JNR assuming $N_j=4$ and the AOAs of the NLJs in between the sampling grid points.}
    \label{fig:case2_RMSE_offgrid}
\end{figure}

The detection performance is shown in Figure \ref{fig:case_2_Pjd}, that clearly confirms the hierarchy 
observed in Figure \ref{p1} with the SC-LRT and SDC-LRT achieving better results than the SPICE-LRT.
Moreover, the presence of an additional NLJ increases the overall JNR and, 
hence, the detection performance. Figures \ref{fig:case_2_EstimationPerf_ongrid}-\ref{fig:case2_hist} 
are related to the classification/estimation performance for the on-grid case and share the same 
parameters as Figures \ref{p2}-\ref{p3} except for $N_j=4$. From the comparisons with respect to
the first operating scenario, it stems that the estimation performance is preserved when the number 
of NLJs changes from $3$ to $4$. In the last two figures, we plot the 
classification histograms as well as the RMS values of the error 
between the actual AOA and the estimated AOA closest to the former. The histograms, reported in 
Figure \ref{fig:case2_hist_offgrid}, show that the SPICE is again more inclined than the proposed
procedure to overestimate the number of NLJs especially for a sampling interval of $1$ degree (note that
also in this case the sum of the occurrences returned by SPICE is less than $1000$, since it 
may estimate a number of jammers greater than $8$). Finally, the comments related to 
Figure \ref{fig:RMSE_offgrid} also hold for Figure \ref{fig:case2_RMSE_offgrid}.

\section{Conclusions}
\label{Sec:Conclusions}
In this paper, we have proposed signal-processing-based radar solutions for the adaptive detection of multiple
NLJs. Specifically, such decision schemes are capable to estimate the number of NLJs illuminating the
radar system and to return their respective AoAs. As a result, the system can draw a picture
of the electromagnetic threats which are active in the radar operating scenario. 
From the design point of view, since the plain MLA leads to intractable optimization problems
from a mathematical point of view (at least to the best of authors' knowledge), we resorted to
a suboptimum approach by developing a systematic framework which relies on cyclic optimizations
and accounts for a sparsity promoting prior at the design stage due to the inherent sparse nature of the problem.
In this context, two adaptive architectures have been devised and assessed using
simulated data. Specifically, the analysis has highlighted that such architectures
can provide reliable detection and estimation performance outperforming their competitor
at least for the considered parameter setting.
Future research tracks might include the design of enhanced fusion strategies 
aimed at handling the artifacts and improving the grid resolution or the extension of the above framework 
to the case of multiple coherent targets. The former issue is part of the current research line. Finally, 
another research line is related to the application of this approach to the new 5G context 
where phased arrays are exploited \cite{5G}.

\section*{Acknowledgements}
This work has been partially supported by NSF under grants No. 1708509 and No. 61971412, as well as by
the EU research project LOCUS No. 871249.

\appendices

\section{Properties of the sparsity-promoting prior }
\label{App:prior}
In this appendix, we describe the properties of $f_d(\bd;\sigma_n^2,q)$ defined by \eqref{eqn:prior} in order to show
that, through $q$, it is possible to {\em tune} its behavior in terms of sparsity promotion. For simplicity,
in what follows we neglect the normalization constant.
It is also important to
remark that $f_d(\bd;\sigma_n^2,q)$ is a suitable modification of the prior introduced in \cite{slim} for the
specific problem at hand (see also the footnote $3$ in Section \ref{Alg:SLIM_known_sigma}).

Let us start by noticing that, given $q=\bar{q}\in(0,1]$, $f_d(\bd;\sigma_n^2,q)$ is continuous and
\begin{align}
\lim_{\|\bd\|^2\rightarrow 0} f_d(\bd;\sigma_n^2,\bar{q}) &<+\infty,
\\
\lim_{\|\bd\|^2\rightarrow +\infty} f_d(\bd;\sigma_n^2,\bar{q}) & =0,
\label{eqn:asymptoticPrior}
\end{align}
where the last equality can be proved by defining $\bD_l=\bD-d_l \boe_l\boe_l^T$, where $\boe_l\in\R^{N\times 1}$ 
is the $l$th vector of the standard basis of $\R^{N\times 1}$, and rewriting the numerator 
of \eqref{eqn:prior} (neglecting, for simplicity, its power) as follows
\begin{align}
&\det(\sigma_n^2\bI+\bV\bD\bV^\dag)=\det(\sigma^2_n\bI+\bV\bD_1\bV^\dag+d_1\bv(\theta_1)\bv(\theta_1)^\dag)\nonumber
\\
&=\det(\sigma_n^2\bI+\bV\bD_1\bV^\dag)\left[1 + d_1\bv^\dag(\theta_1) (\sigma_n^2\bI+\bV\bD_1\bV^\dag)^{-1}  
\bv(\theta_1)\right]
\label{eqn:limNumPrior00}
\\
&=\det(\sigma_n^2\bI+\bV\bD_1\bV^\dag)\left\{1 + d_1\bv^\dag(\theta_1) 
\left[\frac{1}{\sigma_n^{2}}\bI-\frac{1}{(\sigma_n^{2})^2}\bV\bD_1^{1/2}(\bI+\bD_1^{1/2}\bV^\dag\bV\bD_1^{1/2})^{-1}
\bD_1^{1/2}\bV^\dag\right] \bv(\theta_1)\right\}\nonumber
\\
&\leq \det(\sigma_n^2\bI+\bV\bD_1\bV^\dag)\left\{1 + \frac{d_1}{\sigma_n^{2}}\bv^\dag(\theta_1)\bv(\theta_1)\right\}
\nonumber
\\
&=\det(\sigma_n^2\bI+\bV\bD_1\bV^\dag)\left[1 + \frac{d_1}{\sigma_n^{2}}\right],
\label{eqn:limNumPrior}
\end{align}
where the third equality comes from the application of the Woodbury identity \cite{MatrixAnalysis} and the last
inequality is due to the fact that $\bD_1^{1/2}\bV^\dag\bV\bD_1^{1/2}$ is positive semidefinite.
Iterating the above line of reasoning to $\det(\sigma^2_n\bI+\bV\bD_1\bV^\dag)$ and so on by considering
$d_2$, $d_3$, and $d_L$, yields the following inequality
\be
0\leq \frac{\ds[\det(\sigma_n^2\bI + \bV\bD\bV^\dag)]^{K-1}}
{\ds\prod_{l=1}^L \exp\left\{\frac{K}{\bbq}(d_l^{\bbq}-1)\right\}}\leq 
\frac{\ds\left[(\sigma^2_n)^N\prod_{l=1}^{L} \left(1 + \frac{d_l}{\sigma_n^{2}}\right)\right]^{K-1}}
{\ds\prod_{l=1}^L \exp\left\{\frac{K}{\bbq}(d_l^{\bbq}-1)\right\}},
\ee
which allows to apply the {\em Squeeze Theorem} \cite{sohrab2014basic} and \eqref{eqn:asymptoticPrior} 
follows. The above proof also shows that
\be
\forall l=1,\ldots,L: \ \lim_{d_l \rightarrow +\infty} f_d(\bd;\sigma_n^2,\bar{q}) = 0.
\ee
As for the monotonicity of $f_d(\bd;\sigma^2_n,\bar{q})$ with respect to the generic $d_i$, observe that
\be
f_d(\bd;\sigma^2_n,\bar{q})\propto
\frac{\ds[\det(\sigma_n^2\bI + \bV\bD_i\bV^\dag)]^{K-1}}
{\ds\prod_{l=1 \atop l\neq i}^L \exp\left\{\frac{K}{\bbq}(d_l^{\bbq}-1)\right\}}
\underbrace{\frac{\left[1 + d_i\bv^\dag(\theta_i) (\sigma^2_n\bI+\bV\bD_i\bV^\dag)^{-1} \bv(\theta_i)\right]^{K-1}}
{\ds \exp\left\{\frac{K}{\bbq}(d_i^{\bbq}-1)\right\}}}_{p(d_i;\sigma^2_n,\bar{q})}
\ee
and let us study the sign of the first derivative of $p(d_i;\sigma^2_n,\bar{q})$, whose expression is
\be
\frac{\partial}{\partial d_i}p(d_i;\sigma^2_n,\bar{q})=
\frac{A(K-1)(1+A d_i)^{K-2}-(1+A d_i)^{K-1}Kd_i^{\bbq-1}}
{\exp\left\{\frac{K}{\bbq}(d_i^{\bbq}-1)\right\}}
\ee
with $A=\bv^\dag(\theta_1) (\sigma_n^2\bI+\bV\bD_i\bV^\dag)^{-1} \bv(\theta_1) > 0$. Since the denominator
of the above equation is always positive, we focus on the numerator, which can be recast as
\begin{align}
& (1+A d_i)^{K-2}\left[A(K-1)-(1+A d_i)Kd_i^{\bbq-1}\right] \nonumber
\\
&=(1+A d_i)^{K-2}\left[AK(1-d_i^{\bbq})-A-Kd_i^{\bbq-1}\right].
\end{align}
Now, when $d_i \geq 1$, it turns out that $\left[AK(1-d_i^{\bbq})-A-Kd_i^{\bbq-1}\right]<0$ and hence 
$p(d_i;\sigma^2_n,\bbq)$ is strictly decreasing. This behavior is also observed when $0\leq d_i<1$ and $0< A\leq1$.
In the case where $A>1$, there exists a local stationary point in the interval (0, 1). However, the lower bound
on $\sigma^2_n$ guarantees that $A \leq 1$ and that the prior is more oriented to sparsity avoiding the local 
stationary point between $0$ and $1$.

Finally, we consider the limit case $q\rightarrow 0$, which implies that
\be
f_d(\bd;\sigma^2_n,q)\rightarrow 
f_d(\bd;\sigma_n^2)\propto
\frac{\left[ \det(\sigma^2_n\bI+\bV\bD\bV^\dag) \right]^{K-1}}
{\ds \prod_{l=1}^L d_l^{K}},
\ee
where we have used the following well-known result
$\lim_{x\rightarrow 0} \frac{a^x-1}{x}=\log a$.
It is straightforward to show that
$\forall l=1,\ldots,L: \ \lim_{d_l\rightarrow 0}f_d(\bd;\sigma_n^2)=+\infty$.
On the other hand, the limit for large $d_l$ can be computed exploiting \eqref{eqn:limNumPrior00}, which leads 
to the following inequality
\be
\frac{\left[ \det(\sigma^2_n\bI+\bV\bD\bV^\dag) \right]^{K-1}}
{\ds \prod_{l=1}^L d_l^{K}}\leq
\frac{\left[ \det(\sigma^2_n\bI+\bV\bD_i\bV^\dag) \right]^{K-1} 
d_i^{K-1} \left[\bv^\dag(\theta_i) (\sigma^2_n\bI+\bV\bD_i\bV^\dag)^{-1} \bv(\theta_i)\right]^{K-1}}
{\ds \left[\prod_{l=1 \atop l\neq i}^L d_l^{K}\right] d_i^K}.
\ee 
Using the above equation in conjunction with the {\em Squeeze Theorem}, we come up with
$\forall l=1,\ldots,L: \ \lim_{d_l\rightarrow +\infty}f_d(\bd;\sigma_n^2)=0$.
As the last remark, it is not difficult to show that
\be
\forall l=1,\ldots,L: \ \frac{\partial}{\partial d_l} f_d(\bd;\sigma_n^2) < 0, \ d_l>0,
\ee
and, hence, that $f_d(\bd;\sigma_n^2)$ is strictly decreasing with respect to the generic $d_l$.

\section{Cyclic optimization to compute $\widehat{\bd}_q$}
\label{App:compute_d}
Let us consider a preassigned value of $h(q)\in\{1,\ldots,N_{j,\max}\}$ and denote by $\bt$ the vector of integers 
representing the indices of the elements of $\tilde{\bd}_q$ with respect to
$\bd_q^{(n+1)}$ (recall that the former is an ordered copy of the latter). 
Now, we form a vector $\bar{\bd}_q\in\R^{L\times 1}$ such that
\be
\begin{cases}
\bar{\bd}_q(i) = \bd_q^{(n+1)}(i), & \forall k\leq h(q): \ \bt(k)=i,
\\
\bar{\bd}_q(i) = 0, & \mbox{otherwise},
\end{cases}
\ee
namely, the entries of $\bd$, that do not correspond to the selected $h(q)$ peaks, are set to zero.
Assume that an estimate $\bar{\bd}_q^{(n-1)}$ at the $(n-1)$th iteration of the procedure in question is available, 
then, starting from the logarithm of the pdf of $\bZ$ under $H_1$ (namely, the logarithm 
of \eqref{eqn:pdf_Z} under $H_1$), $\forall i\in\bar{\Omega}=\{k\in\N: \ \bar{\bd}_q(k) > 0\}$, we can define 
the following function to be optimized
\begin{align}
g_d(\bar{\bd}_q(i);\bA_{1:i}^{(n-1)})&=-KN\log\pi-K\log\det\left[\bA_{1:i}^{(n-1)} + \bar{\bd}_q(i)\bv(\theta(i))\bv(\theta(i))^\dag\right] \nonumber
\\
&\quad -\tr\left[ \left(\bA_{1:i}^{(n-1)} + \bar{\bd}_q(i)\bv(\theta(i))\bv(\theta(i))^\dag \right)^{-1}\bS \right],
\label{eqn:pdf_Z_00}
\end{align}
where
$
\bA_{1:i}^{(n-1)} = \sigma^2_n\bI + \sum_{k\in\bar{\Omega}\setminus \Omega_{1:i}}\bar{\bd}^{(n-1)}_q(k)\bv(\theta(k))\bv(\theta(k))^\dag
+\bC^{(n)}_i
$
with
$\bC^{(n)}_i=
\ds\sum_{h\in\Omega_{1:i}\setminus \{i\}}\bar{\bd}^{(n)}_q(h)\bv(\theta(h))\bv(\theta(h))^\dag$
and $\Omega_{1:i}=\{k\in\bar{\Omega}: k\leq i\}$. Note that $\bA_{1:i}^{(n-1)}$ is positive definite and can be decomposed as $\bA_{1:i}^{(n-1)}=[\bA_{1:i}^{(n-1)}]^{1/2}[\bA_{1:i}^{(n-1)}]^{1/2}$.
Thus, applying the Woodbury identity \cite{golub1996matrix} and the equality
\be
\det(\bI + \bB_1\bB_2)=\det(\bI + \bB_2\bB_1),
\ee
where $\bB_1\in\C^{N\times M}$ and $\bB_2\in\C^{M\times N}$, equation \eqref{eqn:pdf_Z_00} becomes
\begin{align}
\mbox{\eqref{eqn:pdf_Z_00}}&=-KN\log\pi-K\log\det(\bA_{1:i}^{(n-1)})-K\log\left[1 + \bar{\bd}_q(i)
\bv(\theta(i))^\dag[\bA_{1:i}^{(n-1)}]^{-1}\bv(\theta(i))\right] \nonumber
\\
&\quad -
\tr\left[
\left(
[\bA_{1:i}^{(n-1)}]^{-1} - \bar{\bd}_q(i) \frac{  [\bA_{1:i}^{(n-1)}]^{-1} \bv(\theta(i))\bv(\theta(i))^\dag [\bA_{1:i}^{(n-1)}]^{-1}  }{1+\bar{\bd}_q(i)\bv(\theta(i))^\dag [\bA_{1:i}^{(n-1)}]^{-1} 
\bv(\theta(i))} \right)\bS
\right] \nonumber
\\
&= -KN\log\pi-K\log\det(\bA_{1:i}^{(n-1)})-K\log\left[1 + \bar{\bd}_q(i)\bv(\theta(i))^\dag[\bA_{1:i}^{(n-1)}]^{-1}\bv(\theta(i))\right] \nonumber
\\
&\quad - \tr\left\{ [\bA_{1:i}^{(n-1)}]^{-1}\bS\right\} + \bar{\bd}_q(i)\frac{ \bv(\theta(i))^\dag [\bA_{1:i}^{(n-1)}]^{-1} 
\bS [\bA_{1:i}^{(n-1)}]^{-1} \bv(\theta(i)) }
{1+\bar{\bd}_q(i)\bv(\theta(i))^\dag [\bA_{1:i}^{(n-1)}]^{-1} \bv(\theta(i))}.
\label{eqn:pdf_Z_01}
\end{align}
Setting to zero the first derivative of $g_d(\bar{\bd}_q(i);\bA_{1:i}^{(n-1)})$ with respect to $\bar{\bd}_q(i)$ leads to the following equation
\begin{align}
&\frac{d}{d \bar{\bd}_q(i)}[g_d(\bar{\bd}_q(i);\bA_{1:i}^{(n-1)})]
\\
&=-K\frac{\bv(\theta(i))^\dag[\bA_{1:i}^{(n-1)}]^{-1}\bv(\theta(i))}{1 + \bar{\bd}_q(i)\bv(\theta(i))^\dag[\bA_{1:i}^{(n-1)}]^{-1}\bv(\theta(i))}
+\frac{\bv(\theta(i))^\dag [\bA_{1:i}^{(n-1)}]^{-1} \bS [\bA_{1:i}^{(n-1)}]^{-1} \bv(\theta(i))}
{(1 + \bar{\bd}_q(i)\bv(\theta(i))^\dag [\bA_{1:i}^{(n-1)}]^{-1}\bv(\theta(i)))^2}=0
\\
&\Rightarrow -K\bv(\theta(i))^\dag[\bA_{1:i}^{(n-1)}]^{-1}\bv(\theta(i))-K\bar{\bd}_q(i)\left[\bv(\theta(i))^\dag
[\bA_{1:i}^{(n-1)}]^{-1}\bv(\theta(i))\right]^2 \nonumber
\\
&\quad +\bv(\theta(i))^\dag [\bA_{1:i}^{(n-1)}]^{-1} \bS [\bA_{1:i}^{(n-1)}]^{-1} \bv(\theta(i))=0
\\
&\Rightarrow \widehat{\bar{\bd}}_q(i) = \frac{\bv(\theta(i))^\dag [\bA_{1:i}^{(n-1)}]^{-1} \bS [\bA_{1:i}^{(n-1)}]^{-1} 
\bv(\theta(i))-K\bv(\theta(i))^\dag [\bA_{1:i}^{(n-1)}]^{-1}\bv(\theta(i))}
{K\left[\bv(\theta(i))^\dag [\bA_{1:i}^{(n-1)}]^{-1} \bv(\theta(i))\right]^2}.
\end{align}
Thus, initializing the procedure with $\bar{\bd}_q^{(0)}$ obtained using $\tilde{\bd}_q$ and $\bd_q^{(n+1)}$, 
we can estimate $\bar{\bd}_q$ through the following update rule
\begin{align}
&\forall i\in\bar{\Omega}: \nonumber
\\
&\bar{\bd}^{(n)}_q(i) = \max\left\{\frac{\ds\bv(\theta(i))^\dag \left[\bA_{1:i}^{(n-1)}\right]^{-1} \bS \left[\bA_{1:i}^{(n-1)}\right]^{-1} \bv(\theta(i))
-K\bv(\theta(i))^\dag \left[\bA_{1:i}^{(n-1)}\right]^{-1} \bv(\theta(i))}
{\ds K\left\{\bv(\theta(i))^\dag \left[\bA_{1:i}^{(n-1)}\right]^{-1} \bv(\theta(i))\right\}^2},0\right\}.
\end{align}
Before concluding this appendix an important remark on the convergence of the procedure is in order. 
Specifically, observe that $g_d(\bar{\bd}_q(i);\bA_{1:i}^{(n-1)})$ is continuous, increasing 
when $0\leq\bar{\bd}_q(i)\leq \widehat{\bar{\bd}}_q(i)$, decreasing when 
$\bar{\bd}_q(i) > \widehat{\bar{\bd}}_q(i)$, and
\be
\begin{cases}
\ds\lim_{\bar{\bf d}_q(i)\rightarrow 0^+} g_d(\bar{\bd}_q(i);\bA_{1:i}^{(n-1)}) = C < 0,
\\
\ds\lim_{\bar{\bf d}_q(i)\rightarrow +\infty} g_d(\bar{\bd}_q(i);\bA_{1:i}^{(n-1)}) = -\infty.
\end{cases}
\ee
It follows that there exists a unique global maximum of $g_d(\bar{\bd}_q(i);\bA_{1:i}^{(n-1)})$ 
with respect to $\bar{\bd}_q(i)$ and the iterative procedure gives rise to the 
following increasing sequence
\be
g_d\left(\bar{\bd}^{(0)}_q\right)\leq g_d\left(\bar{\bd}^{(1)}_q\right)\leq \ldots \leq g_d\left(\bar{\bd}^{(n)}_q\right)\leq \ldots,
\label{eqn:sequenceIncreasing}
\ee
where 
\be
g_d\left(\bar{\bd}^{(n)}_q\right) = g_d\left(\bar{\bd}^{(n)}_q(i_1);\bA_{1:i_1}^{(n)}\right) 
\quad \mbox{and} \quad i_1\leq i_2\leq\ldots\leq i_{h(q)}\in\bar{\Omega}.
\ee
In order to prove \eqref{eqn:sequenceIncreasing}, let us note 
that, by construction, the following inequalities hold
\begin{multline}
g_d(\bar{\bd}^{(0)}_q(i_1);\bA_{1:i_1}^{(0)})\leq g_d(\bar{\bd}^{(1)}_q(i_1);\bA_{1:i_1}^{(0)})=g_d(\bar{\bd}^{(0)}_q(i_2);\bA_{1:i_2}^{(0)})
\\
\leq g_d(\bar{\bd}^{(1)}_q(i_2);\bA_{1:i_2}^{(0)})=g_d(\bar{\bd}^{(0)}_q(i_3);\bA_{1:i_3}^{(0)}) 
\leq \ldots \leq g_d(\bar{\bd}^{(1)}_q(i_{h(q)});\bA_{1:i_{h(q)}}^{(0)})
\\
= g_d(\bar{\bd}^{(1)}_q(i_{1});\bA_{1:i_{1}}^{(1)})\leq g_d(\bar{\bd}^{(2)}_q(i_{1});\bA_{1:i_{1}}^{(1)})
\\
\leq\ldots\leq 
g_d(\bar{\bd}^{(n)}_q(i_{h(q)});\bA_{1:i_{h(q)}}^{(n-1)})= g_d(\bar{\bd}^{(n)}_q(i_{1});\bA_{1:i_{1}}^{(n)})\leq\ldots.
\end{multline}
Now, observe that since the function
\begin{multline}
g_d(\bar{\bd}_q)=-KN\log\pi-K\log\det(\sigma^2_n\bI + \bV\diag(\bar{\bd}_q)\bV^\dag)
\\
-\tr\left[ (\sigma^2_n\bI + \bV\diag(\bar{\bd}_q)\bV^\dag)^{-1}\bS \right] , \quad \bar{\bd}_q\in\R^{L\times 1}_+
\end{multline}
is continuous and such that
\be
\begin{cases}
\ds\lim_{\|\bar{\bf d}_q\|\rightarrow 0} g_d(\bar{\bd}_q) = C < 0,
\\
\ds\lim_{\|\bar{\bf d}_q(i)\|\rightarrow +\infty} g_d(\bar{\bd}_q) = -\infty,
\end{cases}
\ee
namely $g_d(\bar{\bd}_q)$ is upper bounded, sequence \eqref{eqn:sequenceIncreasing} does not diverge.
The cyclic optimization, sketched in Algorithm \ref{Alg:refinement_d}, terminates according to a suitable stopping condition based upon
the maximum number of iterations or the estimate variations with respect to the values at the previous iteration.

%
\bibliographystyle{IEEEtran}
\bibliography{group_bib}
%
%

\begin{IEEEbiography}[{\includegraphics[width=1in,height=1.25in,clip,keepaspectratio]{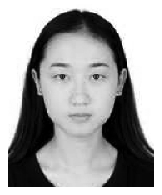}}]{Linjie Yan}
Linjie Yanreceived the B.E. degree in communication engineering from Shandong University of Science and Technology, Shandong, China, in 2016. She is currently working toward the Ph.D. degree in signal and information processing at the Institute of Acoustics,Chinese Academy of Sciences, Beijing, China.
\end{IEEEbiography}

\begin{IEEEbiography}[{\includegraphics[width=1in,height=1.25in,clip,keepaspectratio]{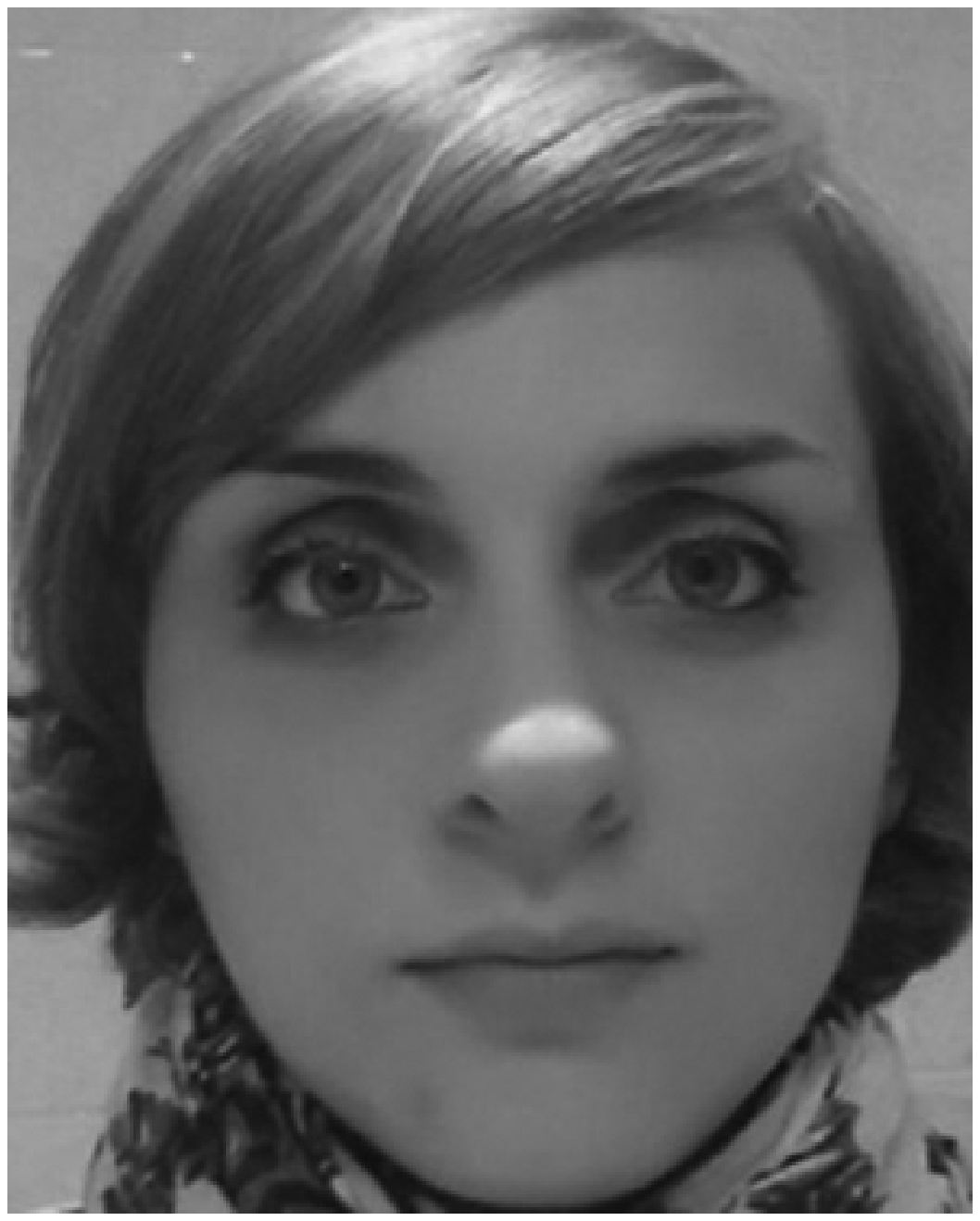}}]{Pia Addabbo}
Pia Addabbo received the B.Sc. and M.Sc. degrees in telecommunication engineering, and the Ph.D. degree in information engineering from the Universit degli Studi del Sannio, Benevento, Italy, in 2005, 2008, and 2012, respectively.,She is a Researcher at the ``Giustino Fortunato'' University, Benevento, Italy. Her research interests include statistical signal processing applied to radar target recognition, global navigation satellite system reflectometry, and hyperspectral unmixing.,Dr. Addabbo is a member of IEEE from 2009 and coauthor of scientific publications in international journals and conferences.
\end{IEEEbiography}

\begin{IEEEbiography}[{\includegraphics[width=1in,height=1.25in,clip,keepaspectratio]{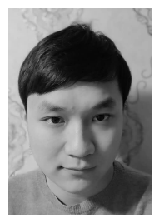}}]{Yuxuan Zhang}
Yuxuan Zhang received the B.E. degree in electronic and information engineering from Harbin Engineering University, Heilongjiang, China, in 2018. He is now studying for the master degree in signal and information processing in Institute of Acoustics, Chinese Academy of Sciences, Beijing, China.
\end{IEEEbiography}

\begin{IEEEbiography}[{\includegraphics[width=1in,height=1.25in,clip,keepaspectratio]{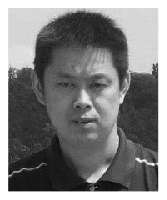}}]{Hao Chengpeng}
Chengpeng Hao(M'08–SM'15) received the B.S. and M.S. degrees in electronic engineering from Beijing Broadcasting Institute, Beijing, China, in 1998 and 2001 respectively,and the Ph.D. degree in signal and information processing from the Institute of Acoustics,Chinese Academy of Sciences, Beijing, China, in 2004.He is currently a Professor with the State Key Laboratory of Information Technology for Autonomous Underwater Vehicles, Chinese Academy of Sciences. He has held a visiting position with the Electrical and Computer Engineering Department, Queens University, Kingston, ON, Canada from July 2013 to July 2014. He authored or coauthored more than 100 journal and conference papers. His research interests are in the fields of statistical signal processing with more emphasis on adaptive sonar and radar signal processing. Dr. Hao is currently serving as an Associate Editor for several international journals,including the IEEE ACCESS,the Signal, Image and Video Processing (Springer), and the Open Electrical and Electronic Engineering Journal. He once served as a Guest Editor for the EURASIP Journal on Advances in Signal Processing for the special issue entitled Advanced Techniques for Radar Signal Processing.
\end{IEEEbiography}

\begin{IEEEbiography}[{\includegraphics[width=1in,height=1.25in,clip,keepaspectratio]{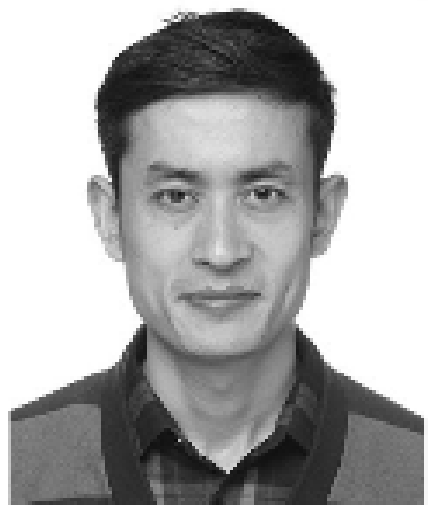}}]{Jun Liu}
Jun  Liu(S'11-M'13-SM'16) received the B.S.degree in mathematics from the Wuhan University of Technology, 
Wuhan, China, in 2006, the M.S. degreein mathematics from Chinese Academy of Sciences,Beijing, China, in 2009, 
and the Ph.D. degree in electrical engineering from Xidian University, Xi'an,China, in 2012.
From July 2012 to December 2012, he was a Postdoctoral Research Associate with the Department of 
Electrical and Computer Engineering, Duke University, Durham, NC, USA. From January 2013 to September 2014, he was a Postdoctoral Research Associate with the Department of Electrical and Computer Engineering, Stevens Institute of Technology,Hoboken, NJ, USA. From October 2014 to March 2018, he was with Xidian University, Xi'an, China. He is currently an Associate Professor with the Department of Electronic Engineering and Information Science, University of Science and Technology of China, Hefei, China. His research interests include statistical signal processing, optimization algorithms, and machine learning. He is currently an Associate Editor for the IEEE SIGNAL PROCESSING LETTERS.
\end{IEEEbiography}

\begin{IEEEbiography}[{\includegraphics[width=1in,height=1.25in,clip,keepaspectratio]{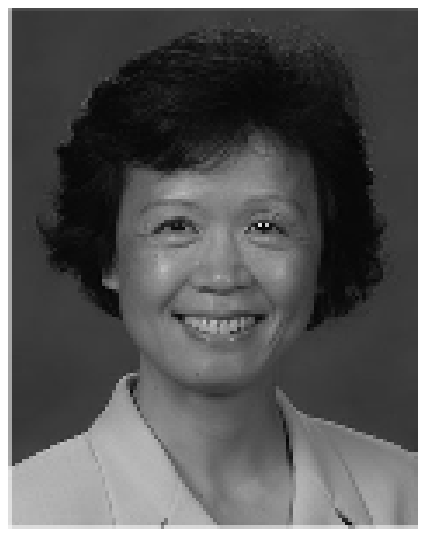}}]{Jian Li}
Jian Li(S'87-M'91-SM'97-F'05) received the M.Sc. and Ph.D. degrees in electrical engineering from The Ohio State University, Columbus, OH,USA, in 1987 and 1991, respectively. She is currently a Professor with the Department of Electrical and Computer Engineering, University of Florida, Gainesville, FL, USA. Her current research interests include spectral estimation,statistical and array signal processing, and their applications to radar, sonar,and biomedical engineering. She has authored Robust Adaptive Beamforming (2005, Wiley), Spectral Analysis: The Missing Data Case (2005, Morgan \& Claypool), MIMO Radar Signal Processing(2009, Wiley), and Waveform Design for Active Sensing Systems-A Computational Approach(2011, Cambridge University Press).Dr. Li is a Fellow of IET. She is also a Fellow of the European Academy of Sciences (Brussels). She was the recipient of the 1994 National Science Foundation Young Investigator Award and the 1996 Office of Naval Research Young Investigator Award. She was an Executive Committee Members of the 2002 and 2016 International Conferences on Acoustics, Speech, and Signal Processing, in Orlando, FL, USA May 2002, and in Shanghai, China, March 2016, respectively. She was an Associate Editor for the IEEE TRANSACTIONS ON SIGNAL PROCESSING from 1999 to 2005, an Associate Editor for the IEEE SIGNAL PROCESSING MAGAZINE from 2003 to 2005, and a member of the Editorial Board of Signal Processing, a publication of the European Association for Signal Processing (EURASIP), from 2005 to 2007. She was a member of the Editorial Board of the IEEE SIGNAL PROCESSING MAGAZINE from 2010 to2012. She is currently a member of the Sensor Array and Multichannel TechnicalCommittee of the IEEE Signal Processing Society. She is a co-author of the paper that has received the M. Barry Carlton Award for the best paper published in the IEEE TRANSACTIONS ON AEROSPACE AND ELECTRONIC SYSTEMS in 2005. She is also a co-author of a paper published in the IEEE TRANSACTIONS ON SIGNAL PROCESSING that has received the Best Paper Award in 2013 from the IEEE Signal Processing Society.
\end{IEEEbiography}

\begin{IEEEbiography}[{\includegraphics[width=1in,height=1.25in,clip,keepaspectratio]{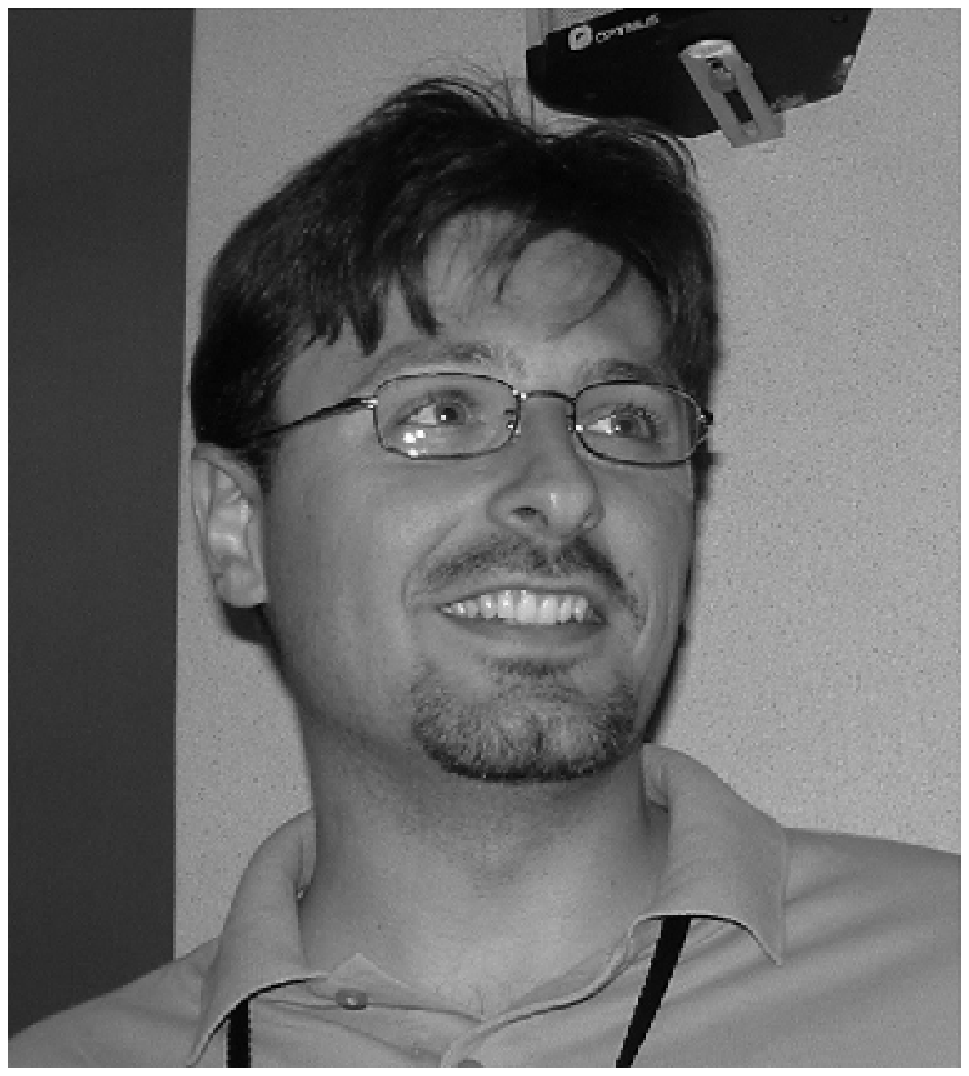}}]{Danilo Orlando}
Danilo Orlando (SM' 13) was born in Gagliano del Capo, Italy, on August 9, 1978. 
He received the Dr. Eng. Degree (with honors) in computer engineering and the
Ph.D. degree (with maximum score) in information engineering, both from the
University of Salento (formerly University of Lecce), Italy, in 2004 and
2008, respectively. From July 2007 to July 2010, he has worked with the
University of Cassino (Italy), engaged in a research project on algorithms
for track-before-detect of multiple targets in uncertain scenarios.
From September to November 2009, he has been visiting scientist at the NATO Undersea
Research Centre (NURC), La Spezia (Italy). From September 2011 to April 2015, he has worked at 
Elettronica SpA engaged as system analyst in the field of Electronic Warfare. 
In May 2015, he joined Università degli Studi ``Niccol\`o Cusano'', where he is currently associate professor.
His main research interests are in the field
of statistical signal processing and image processing with more emphasis on
adaptive detection and tracking of multiple targets in multisensor
scenarios. He has held visiting positions at the department of Avionics
and Systems of ENSICA (now Institut Supérieur de l'Aéronautique et de
l'Espace, ISAE), Toulouse (France) in 2007 and at Chinese Academy of Science, Beijing (China) in 2017-2019. 
He is Senior Member of IEEE; he has served IEEE Transactions on Signal Processing as Senior Area Editor 
and currently is Associate Editor for IEEE Open Journal on Signal Processing,
EURASIP Journal on Advances in Signal Processing, and MDPI Remote Sensing. He is also author or co-author 
of about 110 scientific publications in international journals, conferences, and books.
\end{IEEEbiography}

\vfill
\pagebreak

{\bf Captions of the Figures}
\begin{enumerate}
\item Figure 1:Acquisition procedure of clutter free data for spatial processing.

\item Figure 2: Acquisition procedure of clutter free data for temporal processing.

\item Figure 3: A pictorial representation of the hidden sparse nature of model \eqref{eqn:model_ini} assuming $N_j=2\ll L$.

\item Figure 4: $P_{jd}$ versus JNR for the SC-LRT, the SDC-LRT, and the SPICE-LRT assuming $N_j=3$ and the nominal AOAs for the NLJs.

\item Figure 5: RMS value for the Hausdorff distance, number of missed jammers, and number of ghosts versus JNR assuming $N_j=3$ and the nominal AOAs for the NLJs.

\item Figure 6: Classification histograms for the number of times that the procedures return $1$ jammer,$\ldots$, $6$    jammers assuming $\mbox{JNR}=10$ dB, $N_j=3$, and the nominal AOAs for the NLJs.

\item Figure 7: $P_{jd}$ versus JNR for the SC-LRT, the SDC-LRT, and the SPICE-LRT assuming $N_j=3$, the 
    nominal AOAs for the NLJs, and a JNR variation of $5$ dB during data acquisition.

\item Figure 8: $P_{jd}$ versus JNR for the SC-LRT, the SDC-LRT, and the SPICE-LRT assuming $N_j=3$ and the 
    AOAs of the NLJs in between the sampling grid points.

\item Figure 9: Classification histograms for the number of times that the procedures return $1$ jammer,$\ldots$, $6$
    jammers assuming $\mbox{JNR}=10$ dB, $N_j=3$, and the AOAs of the NLJs in between the sampling grid points.

\item Figure 10: RMS error between the actual AOA of the NLJs and the estimated direction closest to the former 
    versus the JNR assuming $N_j=3$ and the AOAs of the NLJs in between the sampling grid points.

\item Figure 11: $P_{jd}$ versus JNR for the SC-LRT, the SDC-LRT, and the SPICE-LRT assuming $N_j=3$ and the 
    AOAs of the NLJs uniformly generated in a window of size the sampling interval.

\item Figure 12: Classification histograms for the number of times that the procedures return $1$ jammer,$\ldots$, $6$
    jammers assuming $\mbox{JNR}=10$ dB, $N_j=3$, and the AOAs of the NLJs uniformly generated in a 
    window of size the sampling interval.

\item Figure 13: Estimated power (single snapshot) versus search grid angles for three jammers sharing JNR$=30$ dB
    located at: $-10^\circ$, $6^\circ$, and $8^\circ$ subplot (a); $-9.5^\circ$, $-3.5^\circ$, 
    and $8.5^\circ$ subplot (b).

\item Figure 14: $P_{jd}$ versus JNR for the SC-LRT, the SDC-LRT, and the SPICE-LRT assuming $N_j=4$.

\item Figure 15: RMS value for the Hausdorff distance, number of missed jammers, and number of ghosts versus 
    JNR assuming $N_j=4$ and the nominal AOAs for the NLJs.

\item Figure 16: Classification histograms for the number of times that the procedures return $1$ jammer,$\ldots$, $6$
    jammers assuming $\mbox{JNR}=10$ dB, $N_j=4$, and the nominal AOAs for the NLJs.

\item Figure 17: Classification histograms for the number of times that the procedures return $1$ jammer,$\ldots$,$8$
    jammers assuming $\mbox{JNR}=10$ dB, $N_j=4$, and the AOAs of the NLJs in between the sampling grid points.

\item Figure 18: RMS error between the actual AOA of the NLJs and the estimated direction closest to the former 
    versus the JNR assuming $N_j=4$ and the AOAs of the NLJs in between the sampling grid points.

\end{enumerate}

\end{document}